\documentclass[sigconf]{acmart}

\usepackage{bm}
\usepackage{amsthm}
\usepackage{array}
\usepackage{algorithm}
\usepackage{algpseudocode}
\usepackage{amsmath}
\usepackage{amssymb}
\usepackage{balance}
\usepackage{booktabs}
\usepackage{bm}
\usepackage{color}
\usepackage{etex}
\usepackage{graphicx}
\usepackage{multirow}
\usepackage{stmaryrd}
\usepackage{subcaption}
\usepackage{url}
\usepackage{thmtools}
\usepackage{mathtools}
\usepackage{todonotes}
\usepackage{ctable}

\let\bf\textbf

\newcommand{\squishlist}{
	\begin{list}{$\bullet$}
		{ \setlength{\itemsep}{0pt}
			\setlength{\parsep}{3pt}
			\setlength{\topsep}{3pt}
			\setlength{\partopsep}{0pt}
			\setlength{\leftmargin}{1.5em}
			\setlength{\labelwidth}{1em}
			\setlength{\labelsep}{0.5em} } }
	
	\newcommand{\squishlisttwo}{
		\begin{list}{$\bullet$}
			{ \setlength{\itemsep}{0pt}
				\setlength{\parsep}{0pt}
				\setlength{\topsep}{0pt}
				\setlength{\partopsep}{0pt}
				\setlength{\leftmargin}{2em}
				\setlength{\labelwidth}{1.5em}
				\setlength{\labelsep}{0.5em} } }
		
		\newcommand{\squishend}{
	\end{list}
}

\begin{document}
\copyrightyear{2019} 
\acmYear{2019} 
\setcopyright{acmcopyright}
\acmConference[SIGIR '19]{Proceedings of the 42nd International ACM SIGIR Conference on Research and Development in Information Retrieval}{July 21--25, 2019}{Paris, France}
\acmBooktitle{Proceedings of the 42nd International ACM SIGIR Conference on Research and Development in Information Retrieval (SIGIR '19), July 21--25, 2019, Paris, France}
\acmPrice{15.00}
\acmDOI{10.1145/3331184.3331234}
\acmISBN{978-1-4503-6172-9/19/07}

\settopmatter{printacmref=true}
\fancyhead{}
\title{Adversarial Mahalanobis Distance-based Attentive Song Recommender for Automatic Playlist Continuation}

\author{Thanh Tran, Renee Sweeney, Kyumin Lee}
\affiliation{
	\institution{Department of Computer Science \\ Worcester Polytechnic Institute}
	\state{Massachusetts}
	\country{USA}
}
\email{{tdtran, rasweeney, kmlee}@wpi.edu}	

\begin{abstract}
In this paper, we aim to solve the \emph{automatic playlist continuation} (APC) problem by modeling complex interactions among users, playlists, and songs using only their interaction data. Prior methods mainly rely on dot product to account for similarities, which is not ideal as dot product is not metric learning, so it does not convey the important inequality property. Based on this observation, we propose three novel deep learning approaches that utilize Mahalanobis distance. Our first approach uses user-playlist-song interactions, and combines Mahalanobis distance scores between (i) a target user and a target song, and (ii) between a target playlist and the target song to account for both the user's preference and the playlist's theme. Our second approach measures song-song similarities by considering Mahalanobis distance scores between the target song and each member song (i.e., existing song) in the target playlist. The contribution of each distance score is measured by our proposed \emph{memory metric-based attention mechanism}. In the third approach, we fuse the two previous models into a unified model to further enhance their performance. In addition, we adopt and customize \emph{Adversarial Personalized Ranking} (APR) for our three approaches to further improve their robustness and predictive capabilities. Through extensive experiments, we show that our proposed models outperform eight state-of-the-art models in two large-scale real-world datasets. 
\end{abstract}

\maketitle

\section{Introduction}
\label{sec:intro}
The \emph{automatic playlist continuation} (\emph{APC}) problem has received increased attention among researchers following the growth of online music streaming services such as Spotify, Apple Music, SoundCloud, \emph{etc}. Given a user-created playlist of songs, \emph{APC} aims to recommend one or more songs that fit the user's preference and match the playlist's theme. 

Due to the inconsistency of available side information in public music playlist datasets, we first attempt to solve the \emph{APC} problem using only interaction data. \emph{Collaborative filtering} (CF) methods, which encode users, playlists, and songs in lower-dimensional latent spaces, have been widely used \cite{hu2008collaborative,koren2009matrix, he2016fast,he2017neural}. To account for the extra playlist dimension in this work, the term \emph{item} in the context of \emph{APC} will refer to a song, and the term \emph{user} will refer to either a user or playlist, depending on the model. We will also use the term \emph{member song} to denote an existing song within a target playlist. CF solutions to the \emph{APC} problem can be classified into the following three groups:

\smallskip
\noindent\textbf{Group 1: Recommending songs that are directly relevant to user/playlist taste.}
Methods in this group aim to measure the conditional probability of a target item given a target user using user-item interactions. In \emph{APC}, this is either $P(s|u)$ -- the conditional probability of a target song $s$ given a target user $u$ by taking users and songs within their playlists as implicit feedback input --, or $P(s|p)$ -- the conditional probability of a target song $s$ given a target playlist $p$ by utilizing playlist-song interactions. 
Most of the works in this group measure $P(s|u)$\footnote{Measuring $P(s|p)$ is easily obtained by replacing the user latent vector $u$ with the playlist latent vector $p$.} by taking the dot product of the user and song latent vectors \cite{hu2008collaborative,he2016fast,devooght2015dynamic}, denoted by $P(s|u) \propto \overrightarrow{u}^T\cdot\overrightarrow{{s}}$. 
With the recent success of deep learning approaches, researchers proposed neural network-based models \cite{he2017neural,liang2018variational,wu2016collaborative,zhu2019improving}.


Despite their high performance, \emph{Group 1} is limited for the \emph{APC} task. First, although a target song can be very similar to the ones in a target playlist \cite{flexer2008playlist,pohle2005generating}, this association information is ignored. Second, these approaches only measure either $P(s|u)$ or $P(s|p)$, which is sub-optimal. $P(s|u)$ omits the playlist theme, causing the model to recommend the same songs for different playlists, and $P(s|p)$ is not personalized for each user.  

\smallskip
\noindent\textbf{Group 2: Recommending songs that are similar to existing songs in a playlist.}
Methods in this group are based on a principle that similar users prefer the same items (user neighborhood design), or users themselves prefer similar items (item neighborhood design). \emph{ItemKNN}\cite{sarwar2001item}, \emph{SLIM}\cite{ning2011slim}, and \emph{FISM}\cite{kabbur2013fism} solve the \emph{APC} problem by identifying similar users/playlists or similar songs. 
These works are limited in that they give equal weight to each song in a playlist when calculating their similarity to a candidate song. In reality, certain aspects of a member song, such as the genre or artist, may be more important to a user when deciding whether or not to add another song into the playlist. This calls for differing song weights, which are produced by attentive neighborhood-based models.

Recently, \cite{ebesu2018collaborative} proposed a \emph{Collaborative Memory Network} (CMN) that considers both the target user preference on a target item as well as similar users' preferences on that item (i.e., user neighborhood hybrid design). It utilizes a memory network to assign different attentive contributions of neighboring users. However, this approach still does not work well with \emph{APC} datasets due to sparsity issues and less associations among users.



\smallskip
\noindent\textbf{Group 3: Recommending next songs as transitions from the previous songs.}
Methods in this group are called \emph{sequential recommendation} models, which rely on Markov Chains to capture sequential patterns \cite{rendle2010factorizing,wang2015learning}. In the \emph{APC} domain, these methods make recommendations based on the order of songs added to a playlist. 
Deep learning-based sequential methods are able to model even more complex transitional patterns using convolutional neural networks (CNNs) \cite{tang2018personalized} or recurrent neural networks (RNNs) \cite{jing2017neural,donkers2017sequential,hidasi2018recurrent}.
However, \emph{Sequential recommenders} have restrictions in the \emph{APC} domain, namely that playlists are often listened to on shuffle. It means that users typically add songs based on an overarching playlist theme, rather than song transition quality. In addition, added song timestamps may not be available in music datasets.





\begin{figure}
    \centering
    \includegraphics[width=0.4\textwidth]{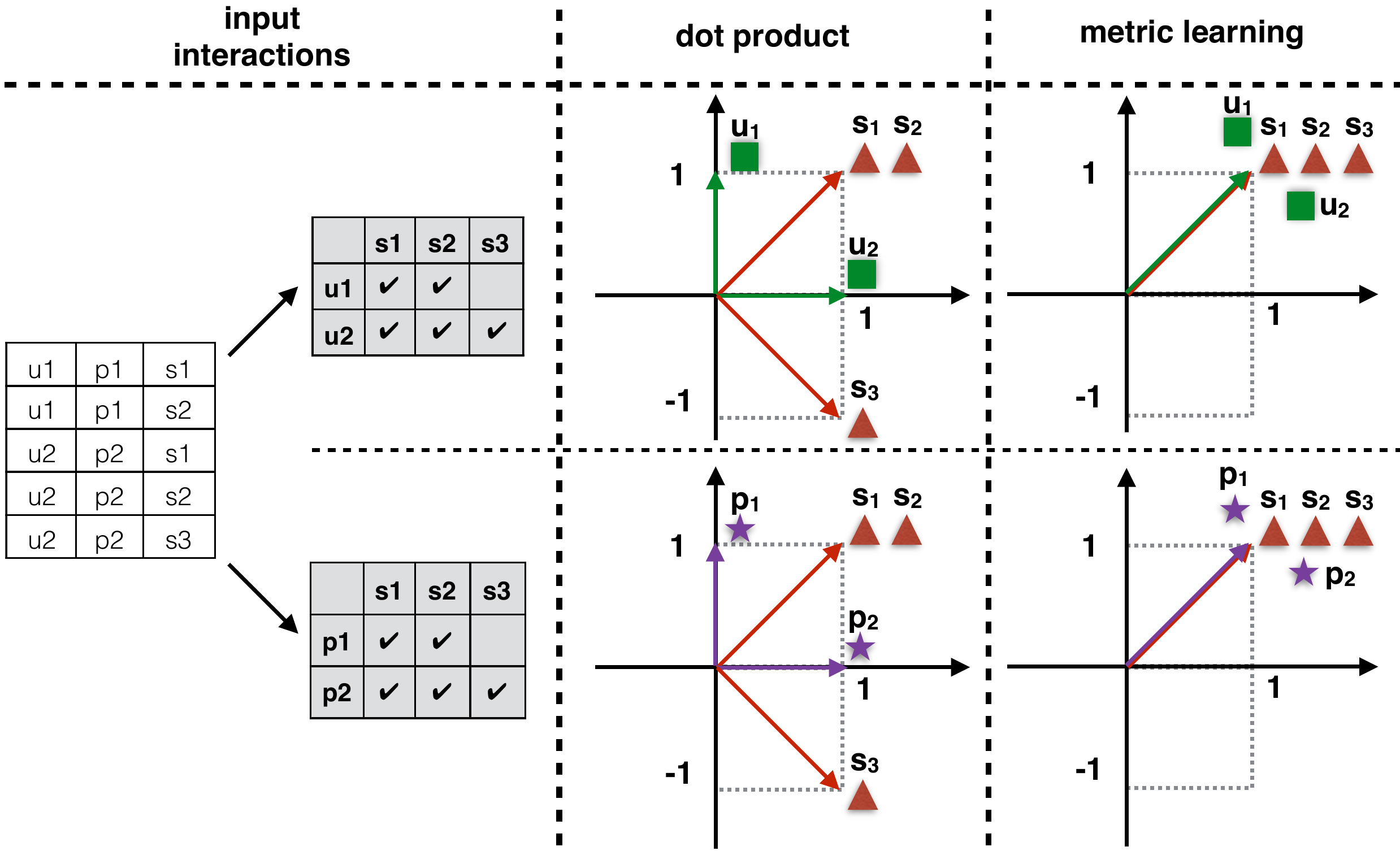}
    \vspace{-10pt}
    \caption{
        Learning with dot product vs. metric learning.
    }
    \label{fig:motivation}
    \vspace{-15pt}
\end{figure}

\smallskip
\noindent\textbf{Motivation.}
A common drawback of the works listed in the three groups above is that they rely on the dot product to measure similarity. Dot product is not metric learning, so it does not convey the crucial inequality property \cite{hsieh2017collaborative,ram2012maximum}, and does not handle differently scaled input variables well. We illustrate the drawback of dot product in a toy example shown in Figure \ref{fig:motivation}\footnote{This Figure is inspired by \cite{hsieh2017collaborative}}, where the latent dimension is size $d=2$. Assume we have two users $u_1$, $u_2$, two playlists $p_1$, $p_2$, and three songs $s_1$, $s_2$, $s_3$. We can see that $p_1$ and $p_2$ (or $u_1$ and $u_2$) are similar (i.e., both liked $s_1$, and $s_2$), suggesting that $s_3$ would be relevant to the playlist $p_1$. Learning with dot product can lead to the following result: $p_1=(0, 1), p_2=(1, 0), s_1=(1, 1), s_2=(1, 1), s_3=(1, -1)$, because $p_1^T s_1=1$, $p_1^T s_2=1$, $p_2^T s_1=1$, $p_2^T s_2=1$, $p_2^T s_3$=1 (same for users $u_1$, $u_2$). However, the dot product between $p_1$ and $s_3$ is -1, so $s_3$ would not be recommended to $p_1$. However, if we use metric learning, it will pull similar users/playlists/songs closer together by using the inequality property. In the example, the distance between $p_1$ and $s_3$ is rescaled to 0, and $s_3$ is now correctly portrayed as a good fit for $p_1$.

There exist several works that adopt metric learning for recommendation. \cite{hsieh2017collaborative} proposed \emph{Collaborative Metric Learning} (CML) which used Euclidean distance to pull positive items closer to a user and push negative items further away. \cite{chen2012playlist,feng2015personalized,he2017translation} also used Euclidean distance but for modeling transitional patterns. However, these metric-based models still fall into either \emph{Group 1} or \emph{Group 3}, inheriting the limitations that we described previously. Furthermore, as Euclidean distance is the primary metric, these models are highly sensitive to the scales of (latent) dimensions/variables.

\smallskip
\noindent\textbf{Our approaches and main contributions.}
According to the literature, Mahalanobis distance\footnote{https://en.wikipedia.org/wiki/Mahalanobis\_distance} \cite{weinberger2006distance,xing2003distance} overcomes the drawback (i.e., high sensitivity) of Euclidean distance. However, Mahalanobis distance has not yet been applied to recommendation with neural network designs.

By overcoming the limitations of existing recommendation models, we propose three novel deep learning approaches in this paper that utilize Mahalanobis distance. Our first approach, \emph{Mahalanobis Distance Based Recommender} (MDR), belongs to \emph{Group 1}. Instead of modeling either $P(s|p)$ or $P(s|u)$, it measures $P(s| u, p)$. To combine both a user' preference and a playlist' theme, \emph{MDR} measures and combines the Mahalanobis distance between the target user and target song, and between the target playlist and target song. Our second approach, \emph{Mahalanobis distance-based Attentive Song Similarity recommender} (MASS), falls into \emph{Group 2}. Unlike the prior works, \emph{MASS} uses Mahalanobis distance to measure similarities between a target song and member songs in the playlist. MASS incorporates our proposed \emph{memory metric-based attention mechanism} that assigns attentive weights to each distance score between the target song and each member song in order to capture different influence levels. Our third approach, \emph{Mahalanobis distance based Attentive Song Recommender} (MASR), combines \emph{MDR} and \emph{MASS} to merge their capabilities. In addition, we incorporate customized \emph{Adversarial Personalized Ranking} \cite{he2018adversarial} into our three models to further improve their robustness.



We summarize our contributions as follows:
\squishlist
    \item We propose three deep learning approaches (\emph{MDR}, \emph{MASS}, and \emph{MASR}) that fully exploit Mahalanobis distance to tackle the \emph{APC} task. As a part of \emph{MASS}, we propose the memory metric-based attention mechanism. 
    \item We improve the robustness of our models by applying \emph{adversarial personalized ranking} and customizing it with a flexible noise magnitude.
    \item We conduct extensive experiments on two large-scale \emph{APC} datasets to show the effectiveness and efficiency of our approaches.
\squishend

\section{Other Related Work}

Music recommendation literature has frequently made use of available metadata such as: lyrics \cite{mcfee2012hypergraph}, tags \cite{mcfee2012hypergraph,liang2015content,jannach2017leveraging,vall2017music,vall2018hybrid}, audio features \cite{mcfee2012hypergraph,wang2014exploration,liang2015content,vall2017music,vall2018hybrid}, audio spectrograms \cite{oramas2017deep,van2013deep,wang2014improving}, song/artist/playlist names \cite{pichl2017improving,jing2017neural,kamehkhosh2017user,aizenberg2012build,teinemaa2018automatic}, and Twitter data \cite{jannach2017leveraging}. Deep learning and hybrid approaches have made significant progress against traditional collaborative filtering music recommenders \cite{wang2014improving,vall2017music,vall2018hybrid,oramas2017deep}.
\cite{wang2014exploration} uses multi-arm bandit reinforcement learning for interactive music recommendation by leveraging novelty and music audio content. \cite{liang2015content} and \cite{van2013deep} perform weighted matrix factorization using latent features pre-trained on a CNN, with song tags and Mel-frequency cepstral coefficients (MFCCs) as input, respectively. 
Unlike these works, our proposed approaches do not require or incorporate side information.

Recently, attention mechanisms have shown their effectiveness
in various machine learning tasks including document classification \cite{yang2016hierarchical}, machine translation \cite{luong2015effective,bahdanau2014neural}, recommendation \cite{ma2018gated,ma2018point}, \emph{etc}. So far, several attention mechanisms are proposed such as: \emph{general attention} \cite{luong2015effective}, \emph{dot} attention \cite{luong2015effective}, \emph{concat} attention \cite{luong2015effective,bahdanau2014neural}, hierarchical attention \cite{yang2016hierarchical}, \emph{scaled dot attention} and \emph{multi-head} attention \cite{vaswani2017attention}, \emph{etc.}. However, to our best of knowledge, most of previously proposed attention mechanisms leveraged dot product for measuring similarities which is not optimal in our Mahalanobis distance-based recommendation approaches because of the difference between dot product space and metric space. Therefore, we propose a memory metric-based attention mechanism for our models' designs.


\section{Problem Definition}


Let ${U} = \{u_1, u_2, u_3, ..., u_{m}$\} denote the set of all users, ${P}$ = \{$p_1$, $p_2$, $p_3$, ..., $p_{n}$\} denote the set of all playlists, ${S}$ = \{$s_1$, $s_2$, $s_3$, ..., $s_{v}$\} denote the set of all songs. Bolded versions of these variables, which we will introduce in the following sections, denote their respective embeddings. \emph{m, n, v} are the number of users, playlists, and songs in a dataset, respectively.
Each user $u_i \in {U}$ has created a set of playlists $T^{(u_i)}$ =\{$p_1$, $p_2$, ..., $p_{|T^{(u_i)}|}$\}, where each playlist $p_j \in T^{(u_i)}$ contains a list of songs $T^{(p_j)}$ =\{$s_1$, $s_2$, ..., $s_{|T^{(p_j)}|}$\}.
Note that $T^{(u_1)} \cup T^{(u_2)} \cup ... \cup T^{(u_m)}$ = \{$p_1, p_2, p_3, ..., p_{n}$\}, $T^{(p_1)} \cup T^{(p_2)} \cup ... \cup T^{(p_n)}$ = \{$s_1, s_2, s_3, ..., s_{v}$\}, and the song order within each playlist is often not available in the dataset.
The \emph{Automatic Playlist Continuity (APC)} problem can then be defined as recommending new songs $s_k \notin T^{(p_j)}$  for each playlist $p_j \in T^{(u_i)}$ created by user $u_i$.

\section{Mahalanobis distance Preliminary}
Given two points $x \in \mathbb{R}^d$ and $y \in \mathbb{R}^d$, the Mahalanobis distance between x and y is defined as:
\vspace{-5pt}
\begin{equation}
\label{equa:Mahalanobis}
d_M(x, y) = \Vert x - y \Vert_M = \sqrt{(x - y)^T M (x-y)}
\end{equation}
where $M \in \mathbb{R}^{d \times d}$ parameterizes the Mahalanobis distance metric to be learned during model training. To ensure that Eq.~(\ref{equa:Mahalanobis}) produces a mathematical metric\footnote{https://en.wikipedia.org/wiki/Metric\_(mathematics)}, $M$ must be symmetric positive semi-definite ($M \succeq 0$). This constraint on $M$ makes the model training process more complicated, so to ease this condition, we rewrite $M = A^T A$ ($A \in \mathbb{R}^{d \times d}$) since $M \succeq 0$. The Mahalanobis distance between two points $d_M(x, y)$ now becomes:
\vspace{-5pt}
\begin{equation}
\label{equa:Mahalanobis-rewrite}
\begin{aligned}
d_M(x, y) = \Vert x - y \Vert_A   &= \sqrt{(x - y)^T A^T A (x-y)}     \\
                                &= \sqrt{\big( A (x - y)\big)^T \big( A (x-y) \big)} \\
                                &= \Vert A(x - y)\Vert_2  = \Vert Ax - Ay \Vert_2
\end{aligned}
\end{equation}
where $\Vert \cdot \Vert_2$ refers to the Euclidean distance.
By rewriting Eq.~(\ref{equa:Mahalanobis}) into Eq.~(\ref{equa:Mahalanobis-rewrite}),
the Mahalanobis distance can now be computed by measuring the Euclidean distance between two linearly transformed points $x \rightarrow Ax$ and $y \rightarrow Ay$. This transformed space encourages the model to learn a more accurate similarity between $x$ and $y$. $d_M(x, y)$ is generalized to basic Euclidean distance $d(x, y)$ when A is the identity matrix. If $A$ in Eq.~(\ref{equa:Mahalanobis-rewrite}) is a diagonal matrix, the objective becomes learning metric $A$ such that different dimensions are assigned different weights. Our experiments show that learning diagonal matrix $A$ generalizes well and produces slightly better performance than if $A$ were a full matrix. Therefore in this paper we focus on only the diagonal case. Also note that when $A$ is diagonal, we can rewrite Eq.~(\ref{equa:Mahalanobis-rewrite}) as:
\begin{equation}
\label{equa:Mahalanobis-rewrite2}
\begin{aligned}
d_M(x, y) = \Vert A(x - y)\Vert_2  =  \Vert diag(A) \odot(x - y)\Vert_2
\end{aligned}
\end{equation}
where $diag(A) \in \mathbb{R}^n$ returns the diagonal of matrix $A$, and $\odot$  denotes the element-wise product. Therefore, we can parameterize $B = diag(A) \in \mathbb{R}^n$ and learn the Mahalanobis distance by simply computing $\Vert B \odot(x - y)\Vert_2$.

In our models' calculations, we will adopt squared Mahalanobis distance, since quadratic form promotes faster learning.

\section{Our proposed models}
In this section, we delve into design elements and parameter estimation of our three proposed models: \emph{Mahalanobis Distance based Recommender (MDR)}, \emph{Mahalanobis distance-based Attentive Song Similarity recommender (MASS)}, and the combined model \emph{Mahalanobis distance based Attentive Song Recommender (MASR)}.

\subsection{Mahalanobis Distance based Recommender (MDR)}
As mentioned in Section \ref{sec:intro}, \emph{MDR} belongs to the \emph{Group 1}. \emph{MDR} takes a target user, a target playlist, and a target song as inputs, and outputs a distance score reflecting the direct relevance of the target song to the target user's music taste and to the target playlist's theme. We will first describe how to measure each of the conditional probabilities -- \emph{P}($s_k | u_i$), \emph{P}($s_k | p_j$), and finally \emph{P}($s_k | u_i, p_j$) -- using Mahalanobis distance. Then we will go over \emph{MDR}'s design.


\vspace{-5pt}
\subsubsection{Measuring P($s_k \vert u_i$)} Given a target user $u_i$, a target playlist $p_j$, a target song $s_k$, and the Mahalanobis distance $d_M(u_i, s_k)$ between $u_i$ and $s_k$, \emph{P}($s_k|u_i$) is measured by:
\vspace{-5pt}
\begin{equation}
\label{equa:prob-u-s}
    \begin{aligned}
P({s_k}|{u_i}) =
                    \frac{\exp(-(d_M^2(\boldsymbol{u_i}, \boldsymbol{s_k}) + \boldsymbol{\beta_{s_k}}))}
                         {\sum_{l} {\exp( -(d_M^2(\boldsymbol{u_i}, \boldsymbol{s_l}) + \boldsymbol{\beta_{s_l}} ) )}}
    \end{aligned}
\end{equation}
where $\beta_{\boldsymbol{s_{k}}}$, $\beta_{\boldsymbol{s_{l}}}$ are bias terms to capture their respective song's overall popularity \cite{koren2009collaborative}. User bias is not included in Eq.(\ref{equa:prob-u-s}) because it is independent of \emph{P}($s_k|u_i$) when varying candidate song $s_k$.
The denominator ${\sum_{l} {\exp( -d_M(\boldsymbol{u_i}, \boldsymbol{s_l}) + \boldsymbol{\beta_{s_l}} )}}$ is a normalization term shared among all candidate songs. 
Thus, \emph{P}($s_k|u_i$) is measured as:
\begin{align}
\label{equa:prob-u-s-aprx}
    P(s_k|u_i) \propto
        -\big(
            d_M^2(\boldsymbol{u_i}, \boldsymbol{s_k}) + \boldsymbol{\beta_{s_k}}
         \big)
\end{align}
Note that training with Bayesian Personalized Ranking (BPR) will only require calculating Eq.~(\ref{equa:prob-u-s-aprx}), since for every pair of observed song ${k^+}$ and unobserved song ${k^-}$ 
, we model the pairwise ranking $P({s_{k^+}}|{u_i}) > P({s_{k^-}}|{u_i})$.
Using Eq.~(\ref{equa:prob-u-s}), this inequality is satisfied only if $d^2_M(\boldsymbol{u_i}, \boldsymbol{s_{k^+}}) + \beta_{\boldsymbol{s_{k^+}}} < d^2_M(\boldsymbol{u_i}, \boldsymbol{s_{k^-}}) + \beta_{\boldsymbol{s_{k^-}}}$, which leads to Eq.~(\ref{equa:prob-u-s-aprx}).

\vspace{-5pt}
\subsubsection{Measuring P($s_k \vert p_j$)} Given a target playlist $p_j$, a target song $s_k$, and the Mahalanobis distance $d_M(p_j, s_k)$ between $p_j$ and $s_k$, P($s_k \vert p_j$) is measured by:

\begin{equation}
\label{equa:prob-p-s}
    \begin{aligned}
P({s_k}|{p_j}) =
            \frac{\exp(-(d_M^2(\boldsymbol{p_j}, \boldsymbol{s_k}) + \boldsymbol{\gamma_{s_k}}))}
                 {\sum_{l} {\exp( -(d_M^2(\boldsymbol{p_j}, \boldsymbol{s_l}) + \boldsymbol{\gamma_{s_l}}))}}
    \end{aligned}
\end{equation}
where $\gamma_{s_k}$ and $\gamma_{s_l}$ are song bias terms. Similar to $P({s_k}|{u_i})$, we shortly measure $P({s_k}|{p_j})$ by:
\begin{align}
\label{equa:prob-p-s-aprx}
    P({s_k}|{p_j}) \propto -(d_M^2(\boldsymbol{p_j}, \boldsymbol{s_k}) + \boldsymbol{\gamma_{s_k}})
\end{align}


\vspace{-5pt}
\subsubsection{Measuring P($s_k \vert u_i, p_j$)}  \emph{P}($s_k|u_i,p_j$) is computed using the Bayesian rule under the assumption that $u_i$ and $p_j$ are conditionally independent given $s_k$:
\begin{equation}
\label{equa:measureP-ups}
\begin{aligned}
    P(s_k|u_i,p_j) &\propto P(u_i|s_k) P(p_j|s_k) P(s_k) \\
                &= \frac{P(s_k|u_i) P(u_i)}{P(s_k)} \frac{P(s_k|p_j) P(p_j)}{P(s_k)} P(s_k) \\
                &\propto P(s_k|u_i) P(s_k|p_j) \frac{1}{P(s_k)} \\
\end{aligned}
\end{equation}
In Eq.~(\ref{equa:measureP-ups}), $P(s_k)$ represents the popularity of target song $s_k$ among the song pool. For simplicity in this paper, we assume that selecting a random candidate song follows a uniform distribution instead of modeling this popularity information. $P(s_k|u_i,p_j)$ then becomes proportional to: $P(s_k|u_i,p_j) \propto  P(s_k|u_i) P(s_k|p_j)$. Using Eq.~(\ref{equa:prob-u-s}, \ref{equa:prob-p-s}), we can approximate $P(s_k|u_i,p_j)$ as follows:
\vspace{-5pt}
\begin{equation}
\label{equa:prob-u-p-s}
\resizebox{0.92\linewidth}{!}{$
\begin{aligned}
& P(s_k|u_i,p_j)  \propto \\
&   \frac{\exp\big( -(d_M^2(\boldsymbol{u_i}, \boldsymbol{s_k}) + \boldsymbol{\beta_{s_k}})
              \big)}
         {\sum_{l} {\exp\big( -(d_M^2(\boldsymbol{u_i}, \boldsymbol{s_l}) + \boldsymbol{\beta_{s_l}})
                        \big)}}
    \times
    \frac{\exp\big(-(d_M^2(\boldsymbol{p_j}, \boldsymbol{s_k}) + \boldsymbol{\gamma{s_k}})
              \big)}
         {\sum_{l} {\exp\big( -(d_M^2(\boldsymbol{p_j}, \boldsymbol{s_l}) + \boldsymbol{\gamma{s_l}})
                        \big)}}
    \\
  =& \frac{\exp \big(-(d_M^2(\boldsymbol{u_i}, \boldsymbol{s_k}) + \boldsymbol{\beta_{s_k}})
                     -(d_M^2(\boldsymbol{p_j}, \boldsymbol{s_k}) + \boldsymbol{\gamma_{s_k}})
                \big)}
         {\sum_{l} {\exp\big( -(d_M^2(\boldsymbol{u_i}, \boldsymbol{s_l}) + \boldsymbol{\beta_{s_l}})
                        \big)}
         \sum_{l\prime} {\exp\big( -(d_M^2(\boldsymbol{p_j}, \boldsymbol{s_{l\prime}}) + \boldsymbol{\gamma_{s_{l\prime}}})
                            \big)}}
\end{aligned}
$}
\end{equation}
Since the denominator of Eq.~(\ref{equa:prob-u-p-s}) is shared by all candidate songs (i.e., normalization term), we can shortly measure $P(s_k|u_i,p_j)$ by:
\vspace{-5pt}
\begin{equation}
\label{equa:prob-u-p-s-aprx}
\begin{aligned}
P(s_k|u_i,p_j)
&\propto
    - \big( d_M^2(\boldsymbol{u_i}, \boldsymbol{s_k}) + d_M^2(\boldsymbol{p_j}, \boldsymbol{s_k}) \big)
    - \big( \boldsymbol{\beta_{s_k}} + \boldsymbol{\gamma_{s_k}} \big)  \\
&=  - \big( d_M^2(\boldsymbol{u_i}, \boldsymbol{s_k}) +
            d_M^2(\boldsymbol{p_j}, \boldsymbol{s_k}) +
            \boldsymbol{\theta_{s_k}}
      \big)
\end{aligned}
\end{equation}

With $P(s_k|u_i,p_j)$ now established in Eq.~(\ref{equa:prob-u-p-s-aprx}), we can move on to our \emph{MDR} model.

\begin{figure}[t]
    \centering
    \includegraphics[width=0.32\textwidth]{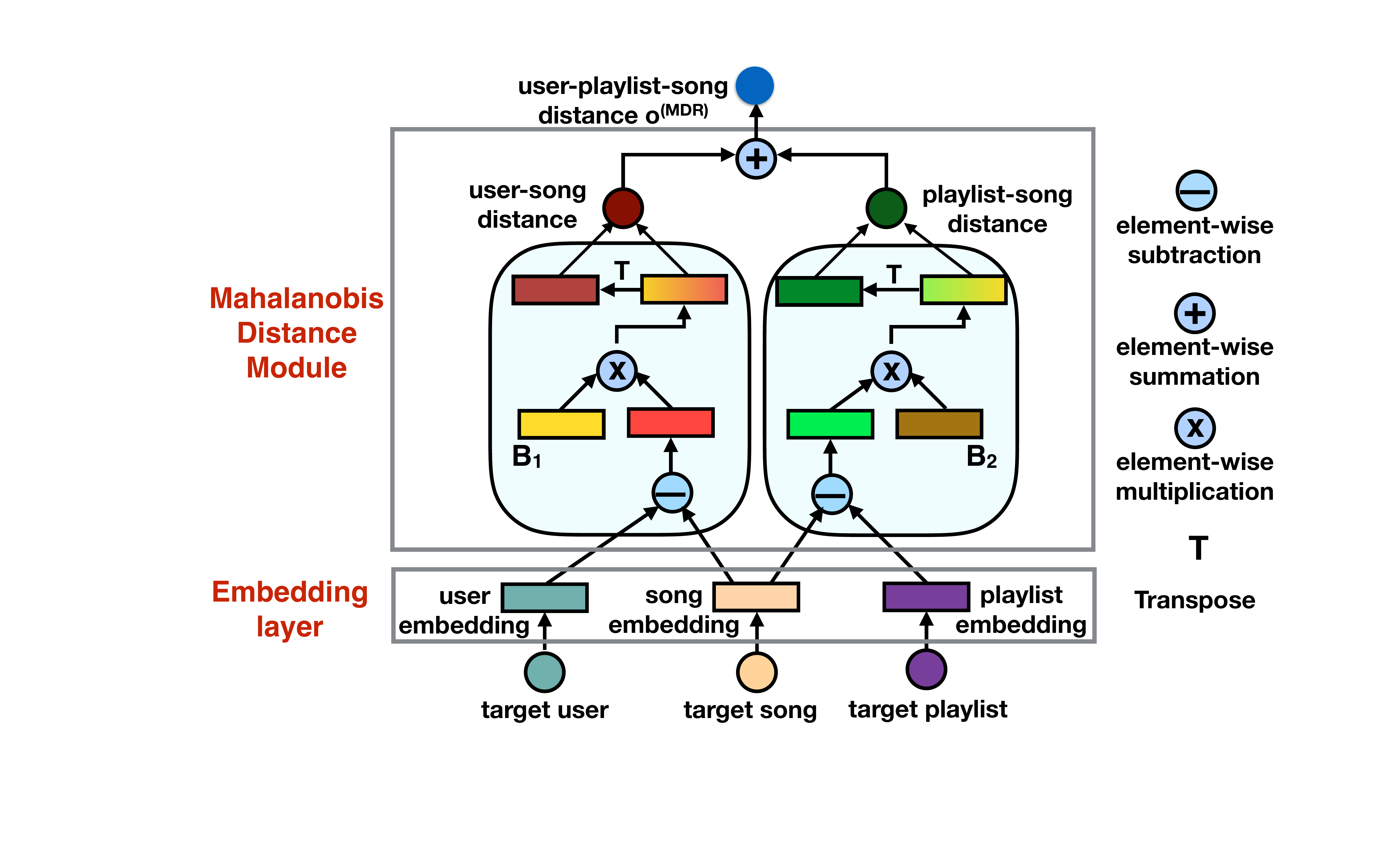}
    \vspace{-5pt}
    \caption{Architecture of our MDR.}
    \label{fig:MahalanobisDistanceRec}
    \vspace{-15pt}
\end{figure}

\vspace{-5pt}
\subsubsection{MDR Design}
The \emph{MDR} architecture is depicted in Figure~\ref{fig:MahalanobisDistanceRec}. It has an Input, Embedding Layer, and Mahalanobis Distance Module.

\noindent\textbf{Input:} \emph{MDR} takes a target user $u_i$ (user ID), a target playlist $p_j$ (playlist ID), and a target song $s_k$ (song ID) as input.

\noindent\textbf{Embedding Layer:} \emph{MDR} maintains three embedding matrices of users, playlists, and songs. By passing user $u_i$, playlist $p_j$, and song $s_k$ through the embedding layer, we obtain their respective embedding vectors $\boldsymbol{u_i} \in \mathbb{R}^d$, $\boldsymbol{p_j} \in \mathbb{R}^d$, and $\boldsymbol{s_k} \in \mathbb{R}^d$, where $d$ is the embedding size.

\noindent\textbf{Mahalanobis Distance Module}: As depicted in Figure~\ref{fig:MahalanobisDistanceRec}, this module outputs a distance score $o^{(MDR)}$ that indicates the relevance of candidate song $s_k$ to both user $u_i$'s music preference and playlist $p_j$'s theme. Intuitively, the lower the distance score is, the more relevant the song is. $o^{(MDR)}(\boldsymbol{u_i}, \boldsymbol{p_j}, \boldsymbol{s_k})$ is computed as follows:
\vspace{-5pt}
\begin{equation}
    \label{equa:mdr-distance}
    {o^{(MDR)}} = o(\boldsymbol{u_i}, \boldsymbol{s_k}) + o(\boldsymbol{p_j}, \boldsymbol{s_k}) + \boldsymbol{\theta_{s_k}}
\end{equation}
where $\boldsymbol{\theta_{s_k}}$ is song $s_k$'s bias, and $o(\boldsymbol{u_i}, \boldsymbol{s_k}), o(\boldsymbol{p_j}, \boldsymbol{s_k})$ are quadratic Mahalanobis distance scores between user $u_i$ and song $s_k$, and between playlist $p_j$ and song $s_k$, shown in the following two equations. $\boldsymbol{B_1} \in \mathbb{R}^d$ and $\boldsymbol{B_2} \in \mathbb{R}^d$ are two metric learning vectors. And,
\vspace{-5pt}
\begin{equation}
\label{equa:mdr-u-s-distance}
\nonumber
\begin{aligned}
    o(\boldsymbol{u_i}, \boldsymbol{s_k}) &= \big( \boldsymbol{B_1} \odot (\boldsymbol{u_i} - \boldsymbol{s_k}) \big)^T
                                             \big( \boldsymbol{B_1} \odot (\boldsymbol{u_i} - \boldsymbol{s_k}) \big) \\
    o(\boldsymbol{p_j}, \boldsymbol{s_k}) &= \big( \boldsymbol{B_2} \odot (\boldsymbol{p_j} - \boldsymbol{s_k}) \big)^T
                                             \big( \boldsymbol{B_2} \odot (\boldsymbol{p_j} - \boldsymbol{s_k}) \big) \\
\end{aligned}
\end{equation}



\subsection{Mahalanobis distance-based Attentive Song Similarity recommender (MASS)}
\begin{figure}[t]
    \centering
    \includegraphics[width=0.36\textwidth]{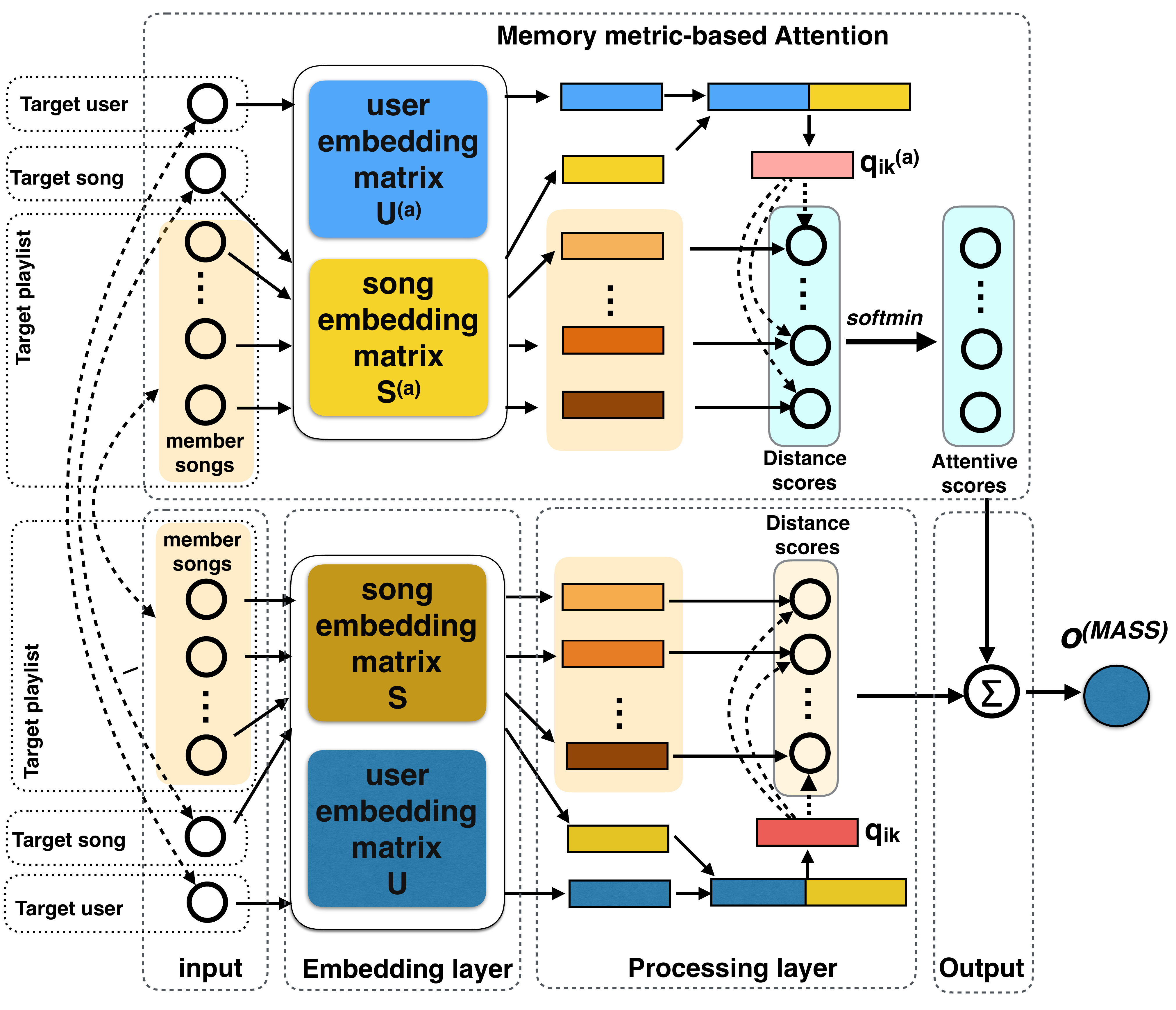}
    \vspace{-10pt}
    \caption{Architecture of our MASS.}
    \label{fig:MASS-model}
    \vspace{-15pt}
\end{figure}

As stated in the Section \ref{sec:intro}, \emph{MASS} belongs to \emph{Group 2}, where it measures attentive similarities between the target song and member songs in the target playlist. An overview of \emph{MASS}'s architecture is depicted in Figure~\ref{fig:MASS-model}. \emph{MASS} has five components: Input, Embedding Layer, Processing Layer, Attention Layer, and Output.
\vspace{-5pt}
\subsubsection{Input:}
\label{sec:input}
The inputs to our \emph{MASS} model include a target user $u_i$, a candidate song $s_k$ for a target playlist $p_j$, and a list of $l$ member songs within the playlist, where $l$ is the number of songs in the largest playlist (i.e., containing the largest number of songs) in the dataset. If a playlist contains less than $l$ songs, we pad the list with \emph{zeroes} until it reaches length $l$. 

\vspace{-5pt}
\subsubsection{Embedding Layer:}
\label{sec:embed-layer}
This layer holds two embedding matrices: a user embedding matrix $\mathbf{U} \in \mathbb{R}^{m \times d}$ and a song embedding matrix $\mathbf{S} \in \mathbb{R}^{v \times d}$. By passing the input target user $u_i$ and target song $s_k$ through these two respective matrices, we obtain their embedding vectors $\boldsymbol{u_i} \in \mathbb{R}^d$ and $\boldsymbol{s_k} \in \mathbb{R}^d$. Similarly, we acquire the embedding vectors for all $l$ member songs in $p_j$, denoted by $\boldsymbol{s_1}, \boldsymbol{s_2}, ..., \boldsymbol{s_l}$.

\vspace{-5pt}
\subsubsection{Processing Layer:} We first need to consolidate $\boldsymbol{u_i}$ and $\boldsymbol{s_k}$. Following widely adopted deep multimodal network designs \cite{srivastava2012multimodal}, we concatenate the two embeddings, and then transform them into a new vector $\boldsymbol{q}_{ik} \in \mathbb{R}^{d}$ via a fully connected layer with weight matrix $\boldsymbol{W}_1 \in \mathbb{R}^{2d \times d}$, bias term $\boldsymbol{b} \in \mathbb{R}$, and activation function \texttt{ReLU}. We formulate this process as follows:
\vspace{-5pt}
\begin{align}
    \label{equa:input-query}
    \boldsymbol{q}_{ik} = \texttt{ReLU} \bigg( \mathbf{W}_1
    \begin{bmatrix}
    \boldsymbol{u}_i \\
    \boldsymbol{s}_k
    \end{bmatrix} + \boldsymbol{b}_1 \bigg)
\end{align}
Note that $\boldsymbol{q}_{ik}$ can be interpreted as a search query in QA systems \cite{xiong2016dynamic,antol2015vqa}. Since we combined the target user $u_i$ with the query song $s_k$ (to add to the user's target playlist), the search query $\boldsymbol{q}_{ik}$ is personalized. The \texttt{ReLU} activation function models a non-linear combination between these two target entities, and was chosen over \emph{sigmoid} or \emph{tanh} due to its encouragement of sparse activations and proven non-saturation \cite{glorot2011deep}, which helps prevent overfitting. 

Next, given the embedding vectors $\boldsymbol{s_1}, \boldsymbol{s_2}, ..., \boldsymbol{s_l}$ of the $l$ member songs in target playlist $p_j$, we approximate the conditional probability \emph{P}$(s_k| u_i, s_1, s_2, ..., s_l)$ by:
\vspace{-5pt}
\begin{align}
P({s}_k| {u}_i, {s}_1, {s}_2, ..., {s}_l)
    \propto
    -\bigg( \sum_{t=1}^{l}  { \boldsymbol{\alpha_{ikt}} d_M^2(\boldsymbol{q_{ik}}, \boldsymbol{s_t)} } +  \boldsymbol{{b}_{s_k}}
    \bigg)
\end{align}
where $d_M(\cdot)$ returns the Mahalanobis distance between two vectors, $b_{s_k}$ is the song bias reflecting its overall popularity, and $\boldsymbol{\alpha}_{ikt}$ is the attention score to weight the contribution of the partial distance between search query $\boldsymbol{q_{ik}}$ and member song $\boldsymbol{s_t}$. We will show how to calculate $d_M^2(\boldsymbol{q_{ik}}, \boldsymbol{s_t})$ below, and $\boldsymbol{\alpha_{ikt}}$ in \emph{Attention Layer} at \ref{sec:att}.

As indicated in Eq.~(\ref{equa:Mahalanobis-rewrite2}), we parameterize $\boldsymbol{B_3} \in \mathbb{R}^d$, which will be learned during the training phase. The Mahalanobis distance between the search query $\boldsymbol{q_{ik}}$ and each member song $\boldsymbol{s_t}$, treating $\boldsymbol{B_3}$ as an edge-weight vector, is measured by:
\begin{equation}
    \label{equa:query-song-dist}
    \resizebox{0.9\linewidth}{!}{$
    \begin{aligned}
        d_M^2(\boldsymbol{q_{ik}}, \boldsymbol{s_t}) &= \big\Vert \boldsymbol{e_{ikt}}^T \boldsymbol{e_{ikt}} \big\Vert_2^2
        \;\;\;\;\; \text{where} \;\;\;\;\;
        \boldsymbol{e_{ikt}} = \boldsymbol{B_3} \odot (\boldsymbol{q_{ik}} - \boldsymbol{s_t})
    \end{aligned}
    $}
\end{equation}


Calculating Eq.~(\ref{equa:query-song-dist}) for every member song $\boldsymbol{s_t}$ yields the following $l$-dimensional vector:

\vspace{-10pt}
\begin{equation}
    \label{equa:distance-scores}
    \begin{aligned}
        \begin{bmatrix}
             d_M^2(\boldsymbol{q}_{ik}, \boldsymbol{s}_1)   \\
             d_M^2(\boldsymbol{q}_{ik}, \boldsymbol{s}_2)   \\
             \dots                                      \\
             d_M^2(\boldsymbol{q}_{ik}, \boldsymbol{s}_l)
        \end{bmatrix} =
        \begin{bmatrix}
            \big\Vert \boldsymbol{e}_{ik1}^T \boldsymbol{e}_{ik1} \big\Vert _2^2  \\
             \big\Vert \boldsymbol{e}_{ik2}^T \boldsymbol{e}_{ik2} \big\Vert _2^2  \\

             \dots                                      \\
             \big\Vert \boldsymbol{e}_{ikl}^T \boldsymbol{e}_{ikl} \big\Vert _2^2  \\
        \end{bmatrix}
    \end{aligned}
\end{equation}
Note that $\boldsymbol{B_3}$ is shared across all Mahalanobis measurement pairs. Now we go into detail of how to calculate the attention weights $\boldsymbol{\alpha_{ikt}}$ using our proposed Attention Layer.

\vspace{-5pt}
\subsubsection{Attention Layer:}
\label{sec:att}
With $l$ distance scores obtained in Eq.~(\ref{equa:distance-scores}), we need to combine them into one distance value to reflect how relevant the target song is \emph{w.r.t} the target playlist's member songs. 
The simplest approach is to follow the well-known item similarity design \cite{ning2011slim,kabbur2013fism} where the same weights are assigned for all $l$ distance scores. This is sub-optimal in our domain because different member song can relate to the target song differently. For example, given a country playlist and a target song of the same genre, the member songs that share the same artist with the target song would be more similar to the target song than the other member songs in the playlist. To address this concern, we propose a novel \emph{memory metric-based attention mechanism} to properly allocate different attentive scores to the distance values in Eq.~(\ref{equa:distance-scores}). Compared to existing attention mechanisms, our attention mechanism maintains its own embedding memory of users and songs (i.e., memory-based property), which can function as an external memory. It also computes attentive scores using Mahalanobis distance (i.e., metric-based property) instead of traditional dot product. Note that the memory-based property is also commonly applied to question-answering in NLP, where memory networks have utilized external memory \cite{sukhbaatar2015end} for better memorization of context information \cite{kumar2016ask,miller2016key}. 
Our attention mechanism has one external memory containing user and song embedding matrices. When the user and song embedding matrices of our attention mechanism are identical to those in the embedding layer, it is the same as looking up the embedding vectors of target users, target songs, and member songs in the embedding layer (Section \ref{sec:embed-layer}). Therefore, using external memory will make room for more flexibility in our models. 

The attention layer features an external user embedding matrix $\mathbf{U^{(a)}} \in \mathbb{R}^{m \times d}$ and external song embedding matrix $\mathbf{S^{(a)}} \in \mathbb{R}^{v \times d}$. Given the following inputs -- a target user $u_i$, a target song $s_k$, and all $l$ member songs in playlist $p_j$ --  by passing them through the corresponding embedding matrices, we obtain the embedding vectors of $u_i$, $s_k$, and all the member songs, denoted as $\boldsymbol{u_i}^{(a)}$, $\boldsymbol{s_k}^{(a)}$, and $\boldsymbol{s_1}^{(a)}, \boldsymbol{s_2}^{(a)}, ..., \boldsymbol{s_l}^{(a)}$, respectively.

We then forge a personalized search query $\boldsymbol{q_{ik}}^{(a)}$ by combining $\boldsymbol{u_i}^{(a)}$ and $\boldsymbol{s_k}^{(a)}$ in a multimodal design as follows:
\vspace{-5pt}
\begin{equation}
\label{equa:att-query}
    \boldsymbol{q_{ik}}^{(a)} = \texttt{ReLU} \bigg( \mathbf{W}_2
    \begin{bmatrix}
    \boldsymbol{u}_i^{(a)} \\
    \boldsymbol{s}_k^{(a)}
    \end{bmatrix} + \boldsymbol{b}_2 \bigg)
\end{equation}
where $\mathbf{W}_2 \in \mathbb{R}^{2d \times d}$ is a weight matrix and $\boldsymbol{b}_2$ is a bias term. Next, we measure the Mahalanobis distance (with an edge weight vector $\boldsymbol{B_4} \in \mathbb{R}^d$) from $\boldsymbol{q_{ik}}^{(a)}$ to a member song's embedding vector $\boldsymbol{s_t}^{(a)}$ where $t \in \overline{1, l}$:
\begin{equation}
\label{equa:att-dist1}
\resizebox{0.9\linewidth}{!}{$
d_M^2( \boldsymbol{q_{ik}}^{(a)}, \boldsymbol{s_t}^{(a)}) = \big\Vert \big(\boldsymbol{e_{ikt}}^{(a)} \big)^T \boldsymbol{e_{ikt}}^{(a)} \big\Vert_2^2
\;\;\; \text{where} \;\;\;
\boldsymbol{e_{ikt}}^{(a)} = \boldsymbol{B_4} \odot \big( \boldsymbol{q_{ik}}^{(a)} - \boldsymbol{s_t^{(a)}} \big)
$}
\end{equation}

Using Eq.~(\ref{equa:att-dist1}), we generate $l$ distance scores between each of $l$ member songs and the candidate song. Then we apply \emph{softmin} on $l$ distance scores in order to obtain the member songs' attentive scores\footnote{Note that attentive scores of padded items are 0.}. Intuitively, the lower the distance between a search query and a member song vector, the higher its contribution level is \emph{w.r.t} the candidate song.


\vspace{-15pt}
\begin{equation}
\label{equa:att-scores}
    \boldsymbol{\alpha_{ikt}} = \frac
    {  \exp{
    \big( -\big\Vert
        \big(\boldsymbol{ e_{ikt}}^{(a)}\big)^T \boldsymbol{e_{ikt}}^{(a)}
    \big\Vert_2^2 \big)} }
    {\sum_{t\prime=1}^{l}
        {
            \exp{ \big( -
                \big\Vert
                \big(\boldsymbol{e_{ikt\prime}}^{(a)}\big)^T \boldsymbol{e_{ikt\prime}}^{(a)} \big\Vert_2^2
                \big)
            }
        }
    }
\end{equation}
\vspace{-10pt}
\subsubsection{Output:}
We output the total attentive distances $o^{(MASS)}$ from the target song $s_k$ to target playlist $p_j$'s existing songs by:
\vspace{-5pt}
\begin{align}
\label{equa:MASS-output}
\mathbf{o^{(MASS)}} = -\bigg( \sum_{t=1}^{l}  { \boldsymbol{\alpha_{ikt}} d_M^2(\boldsymbol{q_{ik}}, \boldsymbol{s_t}) } +  \boldsymbol{b_{s_k}}
\bigg)
\end{align}
where $\boldsymbol{\alpha_{ikt}}$ is the attentive score from Eq.~(\ref{equa:att-scores}), $ d_M(\boldsymbol{q_{ik}}, \boldsymbol{s_t})$ is the personalized Mahalanobis distance between target song $s_k$ and a member song $s_t$ in user $u_i$'s playlist (Eq.~(\ref{equa:distance-scores})), $\boldsymbol{b_{s_k}}$ is the song bias.

\subsection{Mahalanobis distance based Attentive Song Recommender (MASR = MDR + MASS)}
We enhance our performance on the \emph{APC} problem by combining our \emph{MDR} and \emph{MASS} into a \emph{Mahalanobis distance based Attentive Song Recommender (MASR)} model. \emph{MASR} outputs a cumulative distance score from the outputs of \emph{MDR} and \emph{MASS} as follows:
\vspace{-2pt}
\begin{equation}
\label{equa:MASR-output}
\mathbf{o^{(MASR)}} = \alpha \mathbf{o^{(MDR)}} + (1-\alpha) \mathbf{o^{(MASS)}}
\end{equation}
where $\mathbf{o^{(MDR)}}$ is from Eq.~(\ref{equa:mdr-distance}), $\mathbf{o^{(MASS)}}$ is from Eq.~(\ref{equa:MASS-output}), and
$\alpha \in [0,1]$ is a hyperparameter to adjust the contribution levels of \emph{MDR} and \emph{MASS}. $\alpha$ can be tuned using a development dataset. However, in the following experiments, we set $\alpha = 0.5$ to receive equal contribution from \emph{MDR} and \emph{MASS}. 
We pretrain \emph{MDR} and \emph{MASS} first, then fix \emph{MDR} and \emph{MASS}'s parameters in \emph{MASR}. There are two benefits of this design. First, if MASR is learnable with pretrained MDR and MASS initialization, MASR would have too high a computational cost to train. Second, by making \emph{MASR} non-learnable, \emph{MDR} and \emph{MASS} in \emph{MASR} can be trained separately and in parallel, which is more practical and efficient for real-world systems.


\subsection{Parameter Estimation}
\subsubsection{Learning with Bayesian Personalized Ranking (BPR) loss}
We apply BPR loss as an objective function to train our \emph{MDR, MASS, MASR} as follows:
\begin{equation}
\label{equa:bpr-loss}
\resizebox{0.9\linewidth}{!}{$
\mathcal{L}(\mathcal{D} | \Theta) = \operatorname*{argmin}_\Theta \Big( -\sum_{(i, j, k^+, k^-)} \text{log } \sigma(\mathbf{o}_{ijk^-} - \mathbf{o}_{ijk^+}) + \lambda_\Theta \Vert \Theta \Vert^2 \Big)\\
$}
\end{equation}
where $(i, j, k^+, k^-)$ is a quartet of  a target user, a target playlist, a positive song, and a negative song which is randomly sampled. $\sigma(\cdot)$ is the \emph{sigmoid} function; $\mathcal{D}$ denotes all training instances; $\Theta$ are the model's parameters (for instance, $\Theta = \{\mathbf{U}, \mathbf{S}, \mathbf{U^{(a)}}, \mathbf{S^{(a)}}, \mathbf{W_1}, \mathbf{W_2}, \mathbf{B_3},$ $\mathbf{B_4}, \mathbf{b} \}$ in the \emph{MASS} model); $\lambda_\Theta$ is a regularization hyper-parameter; and $\mathbf{o_{ijk}}$ is the output of either \emph{MDR}, \emph{MASS}, or \emph{MASR}, which is measured in Eq.~(\ref{equa:mdr-distance}), (\ref{equa:MASS-output}), and (\ref{equa:MASR-output}), respectively.

\vspace{-5pt}
\subsubsection{Learning with Adversarial Personalized Ranking (APR) loss}
It has been shown in \cite{he2018adversarial} that BPR loss is vulnerable to adversarial noise, and \emph{APR} was proposed to enhance the robustness of a simple matrix factorization model.
In this work, we apply \emph{APR} to further improve the robustness of our \emph{MDR}, \emph{MASS}, and \emph{MASR}. We name our \emph{MDR}, \emph{MASS}, and \emph{MASR} trained with \emph{APR} loss as \emph{AMDR}, \emph{AMASS}, \emph{AMASR} by adding an ``adversarial (A)'' term, respectively. Denote $\delta$ as adversarial noise on the model's parameters $\Theta$. The \emph{BPR} loss from adding adversarial noise $\delta$ to $\Theta$ is defined by:
\vspace{-3pt}
\begin{equation}
\label{equa:bpr-noise-loss}
\mathcal{L}(\mathcal{D} | \hat{\Theta} + \delta) = \operatorname*{argmax}_{\Theta = \hat{\Theta} + \delta} \Big( -\sum_{(i, j, k^+, k^-)} \text{log } \sigma(\mathbf{o}_{ijk^-} - \mathbf{o}_{ijk^+}) \Big)\\
\end{equation}
where $\hat{\Theta}$ is optimized in Eq.~(\ref{equa:bpr-loss}) and fixed as constants in Eq.~(\ref{equa:bpr-noise-loss}). Then, training with \emph{APR} aims to play a minimax game as follows:
\vspace{-3pt}
\begin{equation}
\label{equa:apr-loss}
\operatorname*{arg \; min}_\Theta \operatorname*{max}_{\delta, \Vert \delta \Vert \leq \epsilon s(\hat{\Theta})}
\mathcal{L}(\mathcal{D} | \Theta) + \lambda_\delta \mathcal{L}(\mathcal{D} | \hat{\Theta} + \delta)\\
\end{equation}
where $\epsilon$ is a hyper-parameter to control the magnitude of perturbations $\delta$. In \cite{he2018adversarial}, the authors fixed $\epsilon$ for all the model's parameters, which is not ideal because different parameters can endure different levels of perturbation. If we add too large adversarial noise, the model's performance will downgrade, while adding too small noise does not guarantee more robust models. Hence, we multiply $\epsilon$ with the standard deviation $s(\hat{\Theta})$ of the targeting parameter $\hat{\Theta}$ to provide a more flexible noise magnitude. For instance, the adversarial noise magnitude on parameter $\mathbf{B_3}$ in \emph{AMASS} model is $\epsilon \times s(\mathbf{B_3})$. If the values in $\mathbf{B_3}$ are widely dispersed, they are more vulnerable to attack, so the adversarial noise applied during training must be higher in order to improve robustness. Whereas if the values are centralized, they are already robust, so only a small noise magnitude is needed.

Learning with \emph{APR} follows 4 steps: \bf{Step 1}: unlike \cite{he2018adversarial} where parameters are saved at the last training epoch, which can be over-fitted parameter values (e.g. some thousands of epoches for matrix factorization in \cite{he2018adversarial}), we first learn our models' parameters by minimizing Eq.~(\ref{equa:bpr-loss}) and save the best checkpoint based on evaluating on a development dataset. \bf{Step 2:} with optimal $\hat{\Theta}$ learned in \emph{Step 1}, in Eq.~(\ref{equa:bpr-noise-loss}), we set $\Theta = \hat{\Theta}$ and fix ${\Theta}$ to learn $\delta$. \bf{Step 3:} with optimal $\hat{\delta}$ learned in Eq.~(\ref{equa:bpr-noise-loss}), in Eq.~(\ref{equa:apr-loss}) we set $\delta = \hat{\delta}$ and fix ${\delta}$ to learn new values for $\Theta$. \bf{Step 4}: We repeat \emph{Step 2} and \emph{Step 3} until a maximum number of epochs is reached and save the best checkpoint based on evaluation on a development dataset. Following \cite{he2018adversarial,kurakin2016adversarial}, the update rule for $\delta$ is obtained by using the fast gradient method as follows:
\begin{equation}
\vspace{-5pt}
\label{equa:update-delta}
\delta = \epsilon \times s(\hat{\Theta}) \times
            \frac{\mathbf{\bigtriangledown}_\delta(\mathcal{L}(\mathcal{D} | \hat{\Theta} + \delta))}
            {\big\Vert \mathbf{\bigtriangledown}_\delta(\mathcal{L}(\mathcal{D} | \hat{\Theta} + \delta)) \big\Vert_2}
\end{equation}
Note that update rules of parameters in $\Theta$ can be easily obtained by computing the partial derivative \emph{w.r.t} each parameter in $\Theta$.


\subsection{Time Complexity}

Let $\Omega$ denote the total number of training instances (= $\sum_j{N(p_j)}$ where $N(p_j)$ refers to the number of songs in training playlist $p_j$). $\omega=max(N(p_j))$, $\forall j=\overline{1, n}$ denotes the maximum number of songs in all playlists.
For each forward pass, \emph{MDR} takes $\mathcal{O}(d)$ to measure $\mathbf{o}^{(MDR)}$ (in Eq.~(\ref{equa:mdr-distance})) for a positive training instance, and another forward pass with $\mathcal{O}(d)$ to calculate $\mathbf{o}^{(MDR)}$ for a negative instance.
The backpropagation for updating parameters take the same complexity. Therefore, the time complexity of \emph{MDR} is $\mathcal{O}(\Omega d)$.
Similarly, for each positive training instance, \emph{MASS} takes (i) $\mathcal{O}(2d^2)$ to make each query in Eq.~(\ref{equa:input-query}) and Eq.~(\ref{equa:att-query}); (ii) $\mathcal{O}(\omega d)$ to calculate $\omega$ distance scores from $\omega$ member songs to the target song in Eq.~(\ref{equa:distance-scores}); and (iii) $\mathcal{O}(\omega d)$ to measure attention scores in Eq.~(\ref{equa:att-scores}). Since embedding size $d$ is often small, $\mathcal{O}(\omega d)$ is a dominant term and \emph{MASS}'s time complexity is $\mathcal{O}(\Omega\omega d)$. Hence, both \emph{MDR} and \emph{MASS} scale linearly to the number of training instances and can run very fast, especially with sparse datasets.
When training with \emph{APR}, updating $\delta$ in Eq.~(\ref{equa:update-delta}) with fixed $\hat{\Theta}$ needs one forward and one backward pass. Learning $\Theta$ in Eq.~(\ref{equa:apr-loss}) requires one forward pass to measure $\mathcal{L}(\mathcal{D} | \Theta)$ in Eq.~(\ref{equa:bpr-loss}), one forward pass to measure $\mathcal{L}(\mathcal{D} | \hat{\Theta} + \delta)$ in Eq.~(\ref{equa:bpr-noise-loss}), and one backward pass to update $\Theta$ in Eq.~(\ref{equa:apr-loss}). Hence, time complexity when training with \emph{APR} is $h$ times higher ($h$ is small) compared to training with \emph{BPR} loss.

\section{Empirical study}
\subsection{Datasets}
To evaluate our proposed models and existing baselines, we used two publicly accessible real-world datasets that contain user, playlist, and song information. They are described as follows:
\squishlist
\item 30Music
\cite{turrin201530music}: This is a collection of playlists data retrieved from Internet radio stations through Last.fm\footnote{https://www.last.fm}. It consists of 57K playlists and 466K songs from 15K users.
\item AOTM
 \cite{mcfee2012_dialect}: This dataset was collected from the Art of the Mix\footnote{http://www.artofthemix.org/} playlist database. It consists of 101K playlists and 504K songs from 16K users, spanning from Jan 1998 to June 2011.
\squishend

\begin{table}
	\centering
	\tiny
	\caption{Statistics of datasets.}
	\vspace{-10pt}
	\label{table:datasets}
	\resizebox{0.8\linewidth}{!}{
		\begin{tabular}{lcc}
			\toprule
			Statistics                      & \textbf{30Music}     & \textbf{AOTM} \\
			\midrule
			\# of users                     & 12,336     & 15,835 \\
			\# of playlists                 & 32,140     & 99,903 \\
			\# of songs                     & 276,142    & 504,283 \\
			\# of interactions              & 666,788    & 1,966,795 \\
			avg. \# of playlists per user          & 2.6        & 6.3\\
			avg. \& max \# of songs per playlist      & 18.75 \& 63 &  17.69 \& 58 \\
			Density                         & 0.008\%    & 0.004\% \\
			\bottomrule
		\end{tabular}
	}
	\vspace{-10pt}
\end{table}

For data preprocessing, we removed duplicate songs in playlists. Then we adopted a widely used \emph{k-core} preprocessing step \cite{tran2018regularizing,he2016ups} (with \emph{k-core} = 5), filtering out playlists with less than 5 songs. We also removed users with an extremely large number of playlists, and extremely large playlists (i.e., containing thousands of songs). Since the datasets did not have song order information for playlists (i.e., which song was added to a playlist first, then next, and so on), we randomly shuffled the song order of each playlist and used it in the sequential recommendation baseline models to compare with our models. The two datasets are implicit feedback datasets. The statistics of the preprocessed datasets are presented in Table~\ref{table:datasets}.

\subsection{Baselines}
We compared our proposed models with \textbf{eight} strong state-of-the-art models in the \emph{APC} task. The baselines were trained by using \emph{BPR} loss for a fair comparison:
\squishlist
    \item \textbf{Bayesian Personalized Ranking (MF-BPR)} \cite{rendle2009bpr}: It is a pairwise matrix factorization method for implicit feedback datasets. 

    \item \textbf{Collaborative Metric Learning (CML)} \cite{hsieh2017collaborative}: It is a collaborative metric-based method. It adopted Euclidean distance to measure a user's preference on items. 

    \item \textbf{Neural Collaborative Filtering (NeuMF++)} \cite{he2017neural}: It is a neural network based method that models non-linear user-item interactions. We pretrained two components of NeuMF to obtain its best performance (i.e., NeuMF++).

    \item \textbf{Factored Item Similarity Methods (FISM)} \cite{kabbur2013fism}: It is a item neighborhood-based method. It ranks a candidate song based on its similarity with member songs using dot product. 
    \item \textbf{Collaborative Memory Network (CMN++)} \cite{ebesu2018collaborative}: It is a user-neighborhood based model using a memory network to assign attentive scores for similar users.

    \item \textbf{Personalized Ranking Metric Embedding (PRME)} \cite{feng2015personalized}: It is a sequential recommender that models a personalized first-order Markov behavior using Euclidean distance.

    \item \textbf{Translation-based Recommendation (Transrec)} \cite{he2017translation}: It is one of the best sequential recommendation methods. It models the third order between the user, the previous song, and the next song where the user acts as a translator.

    \item \textbf{Convolutional Sequence Embedding Recommendation} \textbf{\\(Caser)} \cite{tang2018personalized}: It is a CNN based sequential recommendation. It embeds a sequence of recent songs into an ``image'' in time and latent spaces, then learns sequential patterns as local features of the image using different horizontal and vertical filters.
\squishend

We did not compare our models with baselines that performed worse than above listed baselines like \emph{item-KNN}\cite{sarwar2001item}, \emph{SLIM}\cite{ning2011slim}, \emph{etc.}



MF-BPR, CML, and NeuMF++ used only user/playlist-song interaction data to model either users' preferences over songs \emph{P}($s|u$) or playlists' tastes over songs \emph{P}($s|p$). We ran the baselines both ways, and report the best results. Two neighborhood-based baselines utilized neighbor users/playlists (i.e., CMN++) or member songs (i.e., FISM) to recommend the next song based on user/playlist similarities or song similarities (i.e., measure \emph{P}($s|u, s_1, s_2, ..., s_l$) and \emph{P}($s|p, s_1, s_2, ..., s_l$), of which we report the best results). 




\begin{table}
\centering
\tiny
\caption{Performance of the baselines, and our models. The last two lines show the relative improvement of MASR and AMASR compared to the best baseline.}
\vspace{-10pt}
\label{table:PerformanceComparison}
\resizebox{0.45\textwidth}{!}{
    \begin{tabular}{l c@{\hskip0.1pt} cccc}
    \toprule
& \multicolumn{1}{c}{\multirow{2}{*}{Methods}} & \multicolumn{2}{c}{\textbf{30Music}} & \multicolumn{2}{c}{\textbf{AOTM}}
\\
        \cmidrule{3-6}
&                 &  hit@10 & ndcg@10 & hit@10 & ndcg@10 \\
    \midrule
\multicolumn{1}{l@{\hskip0.5pt}|}{(a)}
                    &     MF-BPR        & 0.450 & 0.315 & 0.699 & 0.473
\\
\multicolumn{1}{l@{\hskip0.5pt}|}{(b)}
                    &     CML           & 0.600 & 0.452 & 0.735 & 0.481
\\
\multicolumn{1}{l@{\hskip0.5pt}|}{(c)}
                    &     NeuMF++       & 0.623 & 0.461 & 0.741 & 0.498
\\
\multicolumn{1}{l@{\hskip0.5pt}|}{(d)}
                    &     FISM          & 0.544 & 0.346 & 0.686 & 0.446
\\
\multicolumn{1}{l@{\hskip0.5pt}|}{(e)}
                    &     CMN++         & 0.536 & 0.397 & 0.722 & 0.505
\\
\multicolumn{1}{l@{\hskip0.5pt}|}{(f)}
                    &     PRME          & 0.426 & 0.260 & 0.570 & 0.354
\\
\multicolumn{1}{l@{\hskip0.5pt}|}{(g)}
                    &     Transrec      & 0.570 & 0.417 & 0.710 & 0.450
\\
\multicolumn{1}{l@{\hskip0.5pt}|}{(h)}
                    &     Caser         & 0.458 & 0.289 & 0.681 & 0.448
\\
     \midrule
\multirow{6}{*}{\textbf{Ours  }}
                    &     MDR           & 0.705 & 0.524 & 0.820 & 0.631 \\
                    &     MASS          & 0.670 & 0.500 & 0.834 & 0.639 \\
                    &     MASR          & \bf{0.731} & \bf{0.564} & \bf{0.854} & \bf{0.654} \\
                    \cmidrule{2-6}
                    &     AMDR          & 0.764 & 0.581 & 0.850 & 0.658 \\
                    &     AMASS         & 0.753 & 0.581 & 0.856 & 0.659 \\
                    &     AMASR         & \bf{0.785} & \bf{0.604} & \bf{0.874} & \bf{0.677} \\
    \midrule
\multicolumn{1}{c}{\multirow{3}{*}{\begin{tabular}[c]{@{}c@{}}\bf{Imprv.}\\ \bf{(\%)}\end{tabular}}}
                    &     MASR          & \bf{+17.34} & \bf{+22.34} & \bf{+13.36} & \bf{+28.24}\\
                    &     AMASR         & \bf{+26.00} & \bf{+31.02} & \bf{+17.95} & \bf{+34.19} \\
    \bottomrule
\end{tabular}
}
\vspace{-12pt}
\end{table}

\subsection{Experimental Settings}
\noindent\textbf{Protocol:} We use the widely adopted \emph{leave-one-out} evaluation setting \cite{he2017neural}. Since both the 30Music and AOTM datasets do not contain timestamps of added songs for each playlist, we randomly sample two songs per playlist--one for a positive test sample, and one for a development set to tune hyper-parameters--while the remaining songs in each playlist make up the training set. 
We follow \cite{he2017neural,tran2019signed} and uniformly random sample 100 non-member songs as negative songs, and rank the test song against those negative songs.

\noindent\textbf{Evaluation metrics:} We evaluate the performance of the models with two widely used metrics: Hit Ratio (\emph{hit@N}), and Normalized Discounted Cumulative Gain (\emph{NDCG@N}). The \emph{hit@N} measures whether the test item is in the recommended list or not, while the \emph{NDCG@N} takes into account the position of the \emph{hit} and assigns higher scores to hits at top-rank positions. For the test set, we measure both metrics and report the average scores.

\noindent\textbf{Hyper-parameters settings:} Models are trained with the \emph{Adam} optimizer with learning rates from \{0.001, 0.0001\}, regularization terms $\lambda_\Theta$ from \{0, 0.1, 0.01, 0.001, 0.0001\}, and embedding sizes from \{8, 16, 32, 64\}. The maximum number of epochs is 50, and the batch size is 256. The number of hops in \emph{CMN++} are selected from \{1, 2, 3, 4\}. In \emph{NeuMF++}, the number of MLP layers are selected from \{1, 2, 3\}. The number of negative samples per one positive instance is 4, similar to \cite{he2017neural}. The Markov order $L$ in \emph{Caser} is selected from \{4, 5, 6, 7, 8, 9, 10\}. For \emph{APR} training, the number of \emph{APR} training epochs is 50, the noise magnitude $\epsilon$ is selected from \{0.5, 1.0\}, and the adversarial regularization $\lambda_\delta$ is set to 1, as suggested in \cite{he2018adversarial}. Adversarial noise is added only in training process, and are initialized as \emph{zero}. All hyper-parameters are tuned by using the development set. Our source code is available at \emph{https://github.com/thanhdtran/MASR.git}.


\begin{table}[t]
    \centering
    \tiny
    \caption{Performance of variants of our MDR and MASS. RI indicates relative average improvement over the corresponding method.}
    \vspace{-10pt}
    \label{table:Effect-playlist}
    \resizebox{0.45\textwidth}{!}{
        \begin{tabular}{l p{8pt} p{15pt} p{8pt} p{15pt} c}
        \toprule
\multicolumn{1}{c}{\multirow{2}{*}{Methods}} & \multicolumn{2}{c}{\textbf{30Music}} & \multicolumn{2}{c}{\textbf{AOTM}} & \multicolumn{1}{c}{\multirow{2}{*}{\bf{RI(\%)}}} \\
        \cmidrule{2-5}
                        & hit@10 & ndcg@10 & hit@10 & ndcg@10 & \\
    \midrule

MDR\_us      & 0.684 & 0.500 & 0.815 & 0.594 & \multicolumn{1}{|c}{\bf{+3.68}}\\
MDR\_ps      & 0.654 & 0.476 & 0.746 & 0.547 & \multicolumn{1}{|c}{\bf{+10.79}}\\
\cmidrule{2-6}
MDR\_ups (i.e., MDR)      & \bf{0.705} & \bf{0.524} & \bf{0.818} & \bf{0.613} \\
    \midrule
MASS\_ups    & 0.651 & 0.479 & 0.789 & 0.581 & \multicolumn{1}{|c}{\bf{+4.12}}\\
MASS\_ps     & 0.621 & 0.450 & 0.764 & 0.523 & \multicolumn{1}{|c}{\bf{+10.82}}\\
    \cmidrule{2-6}
MASS\_us (i.e., MASS)      & \bf{0.670} & \bf{0.500} & \bf{0.820} & \bf{0.631} &\\
    \bottomrule
        \end{tabular}
}
\vspace{-12pt}
\end{table}
\subsection{Experimental Results}
\label{sec:results}
%

\subsubsection{Performance comparison}
Table \ref{table:PerformanceComparison} shows the performance of our proposed models and baselines on each dataset. \emph{MDR} and baselines (a)-(c) are in \emph{Group 1}, but \emph{MDR} shows much better performance compared to the (a)-(c) baselines, improving at least 11.14\% \emph{hit@10} and 18.81\% \emph{NDCG@10} on average. \emph{CML} simply adopts Euclidean distance between users/playlists and positive songs, but has nearly equal performance with NeuMF++, which utilizes a neural network to learn non-linear relationships between users/playlists and songs. This result shows the effectiveness of using metric learning over dot product in recommendation. \emph{MDR} outperforms \emph{CML} by 19.04\% on average. This confirms the effectiveness of Mahalanobis distance over Euclidian distance for recommendation.

\emph{MASS} outperforms both \emph{FISM} and \emph{CMN++}, improving \emph{hit@10} by 18.4\%, and \emph{NDCG@10} by 25.5\% on average. This is because \emph{FISM} does not consider the attentive contribution of different neighbors. Even though \emph{CMN++} can assign attention scores for different user/playlist neighbors, it bears the flaws of \emph{Group 1} by considering only either neighbor users or neighbor playlists. More importantly, \emph{MASS} uses a novel attentive metric design, while dot product is utilized in \emph{FISM} and \emph{CMN++}. Sequential models, (f)-(h) baselines, do not work well. In particular, \emph{MASS} outperforms the (f)-(h) baselines, improving 24.6\% on average compared to the best model in (f)-(h).

\emph{MASR} outperforms both \emph{MDR} and \emph{MASS}, indicating the effectiveness of fusing them into one model. Particularly, \emph{MASR} improves \emph{MDR} by 5.0\%, and \emph{MASS} by 6.7\% on average. Performances of \emph{MDR, MASS, MASR} are boosted when adopting \emph{APR} loss with a \emph{flexible} noise magnitude. \emph{AMDR} improves \emph{MDR} by 7.7\%, \emph{AMASS} improves \emph{MASS} by 9.4\%, and \emph{AMASR} improves \emph{MASR} by 5.8\%. We also compare our \emph{flexible} noise magnitude with a fixed noise magnitude used in \cite{he2018adversarial} by varying the fixed noise magnitude in \{0.5, 1.0\} and setting $\lambda_\delta = 1$. We observe that \emph{APR} with a \emph{flexible} noise magnitude performs better with an average improvement of 7.53\%.

Next, we build variants of our \emph{MDR} and \emph{MASS} models by removing either playlist or user embeddings, or using both of them. Table \ref{table:Effect-playlist} presents an ablation study of exploiting playlist embeddings. \emph{MDR\_us} is the \emph{MDR} that uses only user-song interactions (i.e., ignore playlist-song distance $o(p_j, s_k)$ in Eq.~(\ref{equa:mdr-distance})). \emph{MDR\_ps} is the \emph{MDR} that uses only playlist-song interactions (i.e., ignores user-song distance $o(u_i, s_k)$ in Eq.~(\ref{equa:mdr-distance})). \emph{MDR\_ups} is our proposed \emph{MDR} model. Similarly, \emph{MASS\_ups} is the \emph{MASS} model but considers both user-song distances and playlist-song distances in its design. The \emph{Embedding Layer} and \emph{Attention Layer} of \emph{MASS\_ups} have additional playlist embedding matrices $\mathbf{P} \in \mathbb{R}^{n\times d}$ and $\mathbf{P^{(a)}} \in \mathbb{R}^{n\times d}$, respectively. \emph{MASS\_ps} is the \emph{MASS} model that replaces user embeddings with playlist embeddings. \emph{MASS\_us} is our proposed \emph{MASS} model.

\emph{MDR} (i.e., \emph{MDR\_ups})  outperforms its derived forms (\emph{MDR\_us} and \emph{MDR\_ps}), improving by 3.7$\sim$10.8\% on average. This result shows the effectiveness of modeling both users' preferences and playlists' themes in \emph{MDR} design. \emph{MASS} (i.e., \emph{MASS\_us}) outperforms its two variants (\emph{MASS\_ups} and \emph{MASS\_ps}), improving \emph{MASS\_ups} by 3.7\%, and \emph{MASS\_ps} by 10.8\% on average. It makes sense that using additional playlist embeddings in \emph{MASS\_ups} is redundant since the member songs have already conveyed the playlist's theme, and ignoring user embeddings in \emph{MASS\_ps} neglects user preferences.

\begin{figure}[t]
    \includegraphics[width=0.46\textwidth]{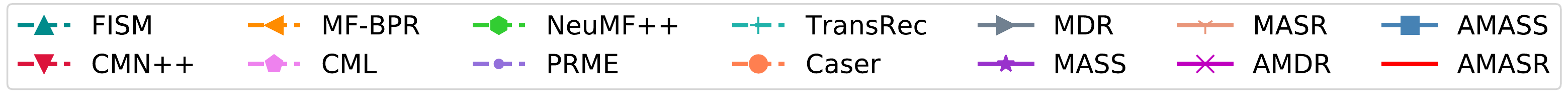}
    \centering
	\begin{subfigure}{0.23\textwidth}
		\centering
		\includegraphics[width=0.96\textwidth]{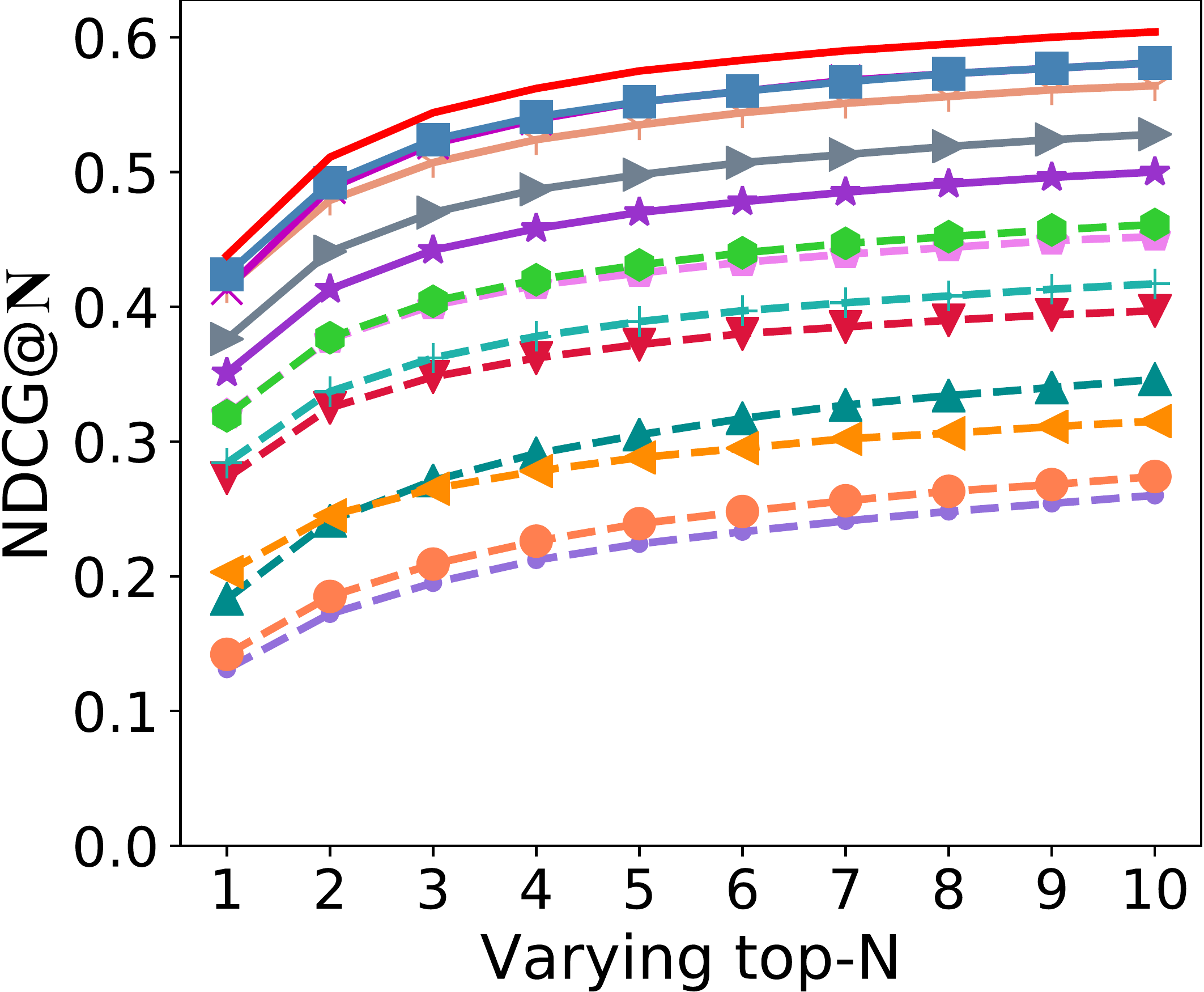}
		\vspace{-5pt}
		\caption{30Music.}
		\label{fig:30music-hit-varytopN}
	\end{subfigure}
	\hspace{-5pt}
	\begin{subfigure}{0.23\textwidth}
		\centering
		\includegraphics[width=0.96\textwidth]{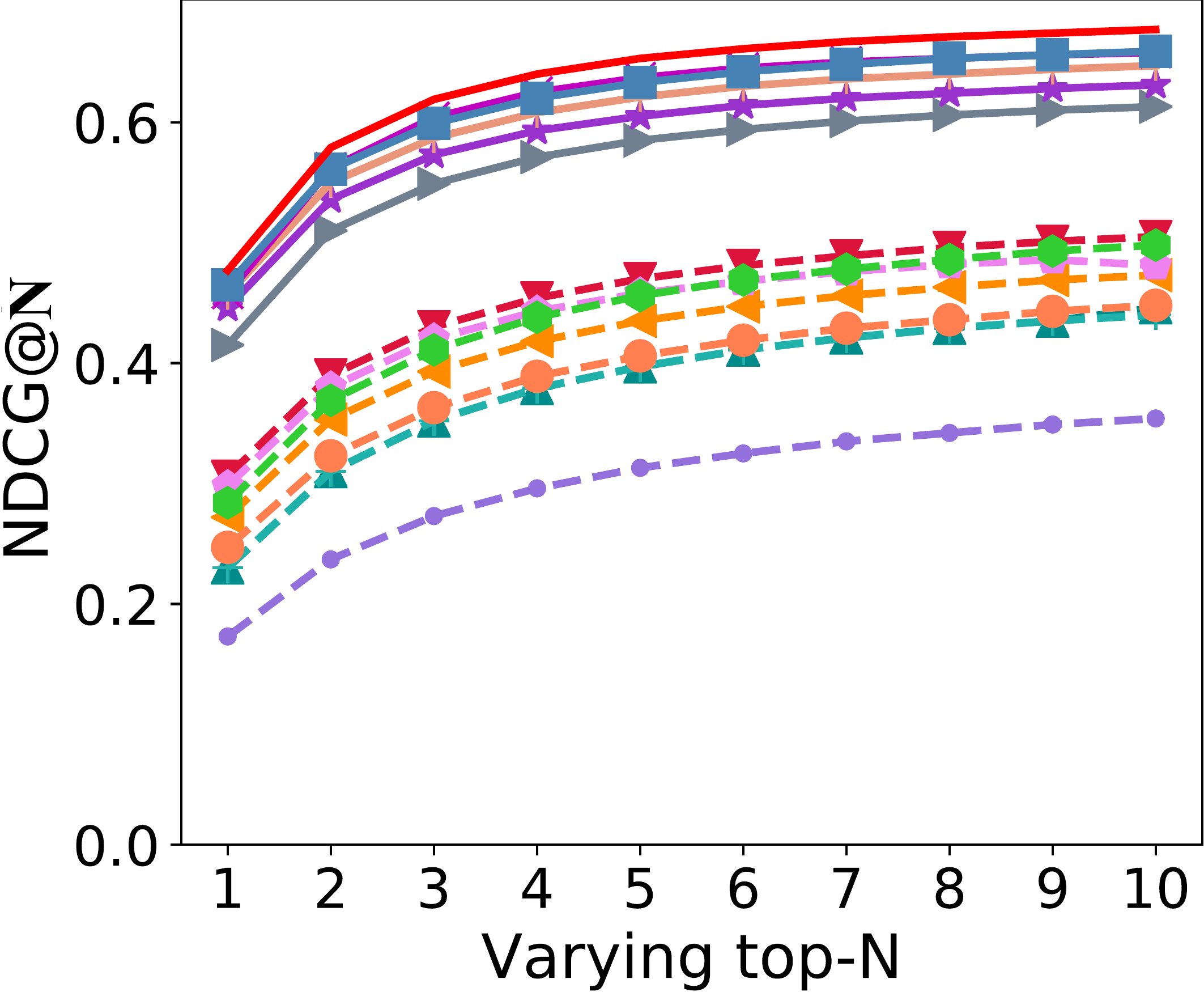}
		\vspace{-5pt}
		\caption{AOTM.}
		\label{fig:aotm-hit-varytopN}
	\end{subfigure}
    \vspace{-10pt}
    \caption{Performance of our models and the baselines when varying \emph{N} (or \emph{top-N} recommendation list) from [1, 10].}
    \label{fig:varytopN-performance}
    \vspace{-10pt}
\end{figure}

\begin{figure}[t]
    \centering
 	\begin{subfigure}{0.46\textwidth}
		\includegraphics[width=0.46\textwidth]{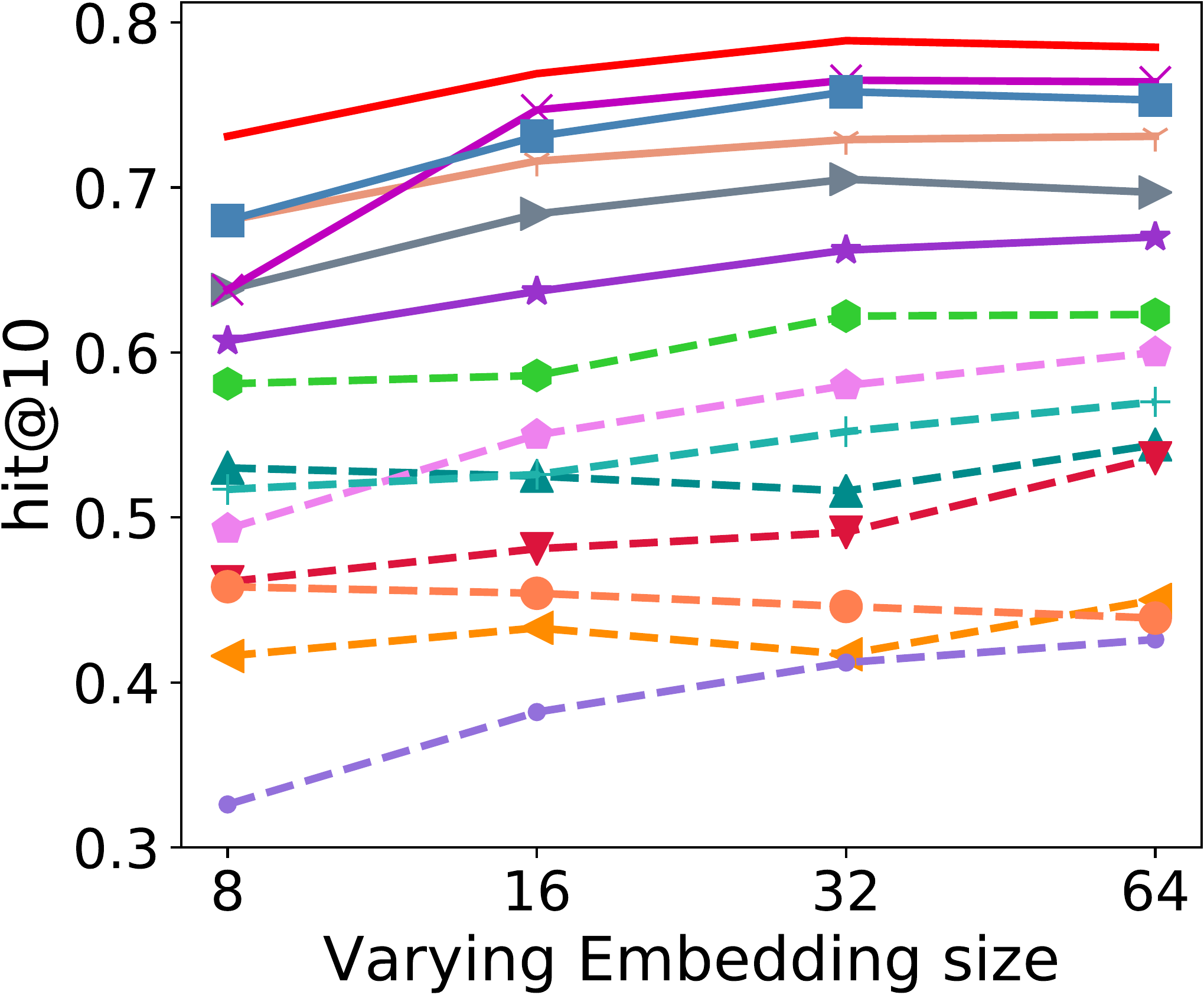}
		\hspace{5pt}
		\includegraphics[width=0.46\textwidth]{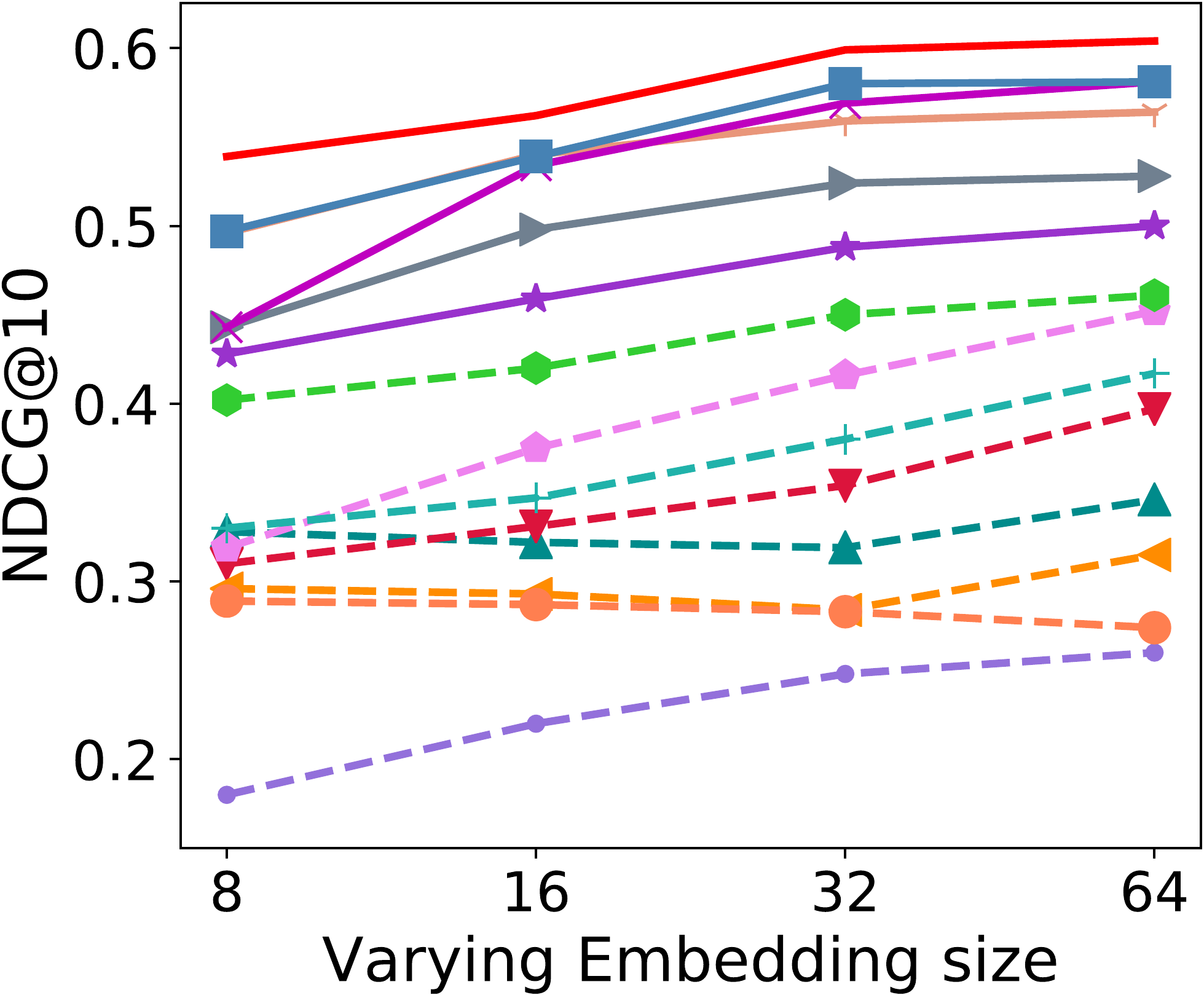}
 	\end{subfigure}
    \vspace{-10pt}
    \caption{Performance of all models when varying the embedding size \emph{d} from \{8, 16, 32, 64\} in 30Music dataset.}
    \label{fig:varyembedding-performance}
    \vspace{-10pt}
\end{figure}

\vspace{-5pt}
\subsubsection{Varying top-N recommendation list and embedding size}
Figure~\ref{fig:varytopN-performance} shows performances of all models when varying \emph{top-N} recommendation from 1 to 10. We see that all models gain higher results when increasing \emph{top-N}, and all our proposed models outperform all baselines across all \emph{top-N} values. On average, \emph{MASR} improves 26.3\%, and \emph{AMASR} improves 33.9\% over the best baseline's performance.

Figure \ref{fig:varyembedding-performance}\footnote{Figure \ref{fig:varyembedding-performance} shares the same legend with Figure \ref{fig:varytopN-performance} for saving space.} shows all models' performances when varying the embedding size $d$ from \{8, 16, 32, 64\} for the \emph{30Music} dataset. Note that the \emph{AOTM} dataset also shows similar results but is omitted due to the space limitations. We observe that most models tend to have increased performance when increasing embedding size. \emph{AMDR} does not improve \emph{MDR} when $d=8$ but does so when increasing $d$. This phenomenon was also reported in \cite{he2018adversarial} because when $d=8$, \emph{MDR} is too simple and has a small number of parameters, which is far from overfitting the data and not very vulnerable to adversarial noise. However, for more complicated models like \emph{MASS} and \emph{MASR}, even with a small embedding size $d=8$, \emph{APR} shows its effectiveness in making the models more robust, and leads to an improvement of \emph{AMASS} by 12.0\% over \emph{MASS}, and an improvement of \emph{AMASR} by 7.5\% over \emph{MASR}. The improvements of \emph{AMDR, AMASS, AMASR} over their corresponding base models are higher for larger $d$ due to the increase of model complexity.

\begin{table}[t]
    \centering
    \tiny
    \caption{Performance of MASS using various attention mechanisms.}
    \vspace{-10pt}
    \label{table:attentionComparision}
    \resizebox{0.46\textwidth}{!}{
        \begin{tabular}{lccccr}
        \toprule
\multicolumn{1}{c}{\multirow{2}{*}{Attention Types}} & \multicolumn{2}{c}{\textbf{30Music}} & \multicolumn{2}{c}{\textbf{AOTM}} & \multicolumn{1}{c}{\multirow{2}{*}{\bf{RI(\%)}}} \\
        \cmidrule{2-5}
                        & hit@10 & ndcg@10 & hit@10 & ndcg@10 & \\
    \midrule

non-mem + dot      & 0.630 & 0.454 & 0.785 & 0.574 & \multicolumn{1}{|c}{\bf{+8.51}}\\
non-mem + metric   & 0.660 & 0.490 & 0.803 & 0.601 & \multicolumn{1}{|c}{\bf{+3.43}}\\
mem + dot          & 0.659 & 0.475 & 0.791 & 0.585 & \multicolumn{1}{|c}{\bf{+5.40}}\\
\midrule
\bf{mem + metric}   & \bf{0.670} & \bf{0.500} & \bf{0.834} & \bf{0.639} & \\
    \bottomrule
        \end{tabular}
    }
    \vspace{-10pt}
\end{table}
\begin{figure}[t]
    \centering
 	\begin{subfigure}{0.11\textwidth}
		\includegraphics[width=\textwidth]{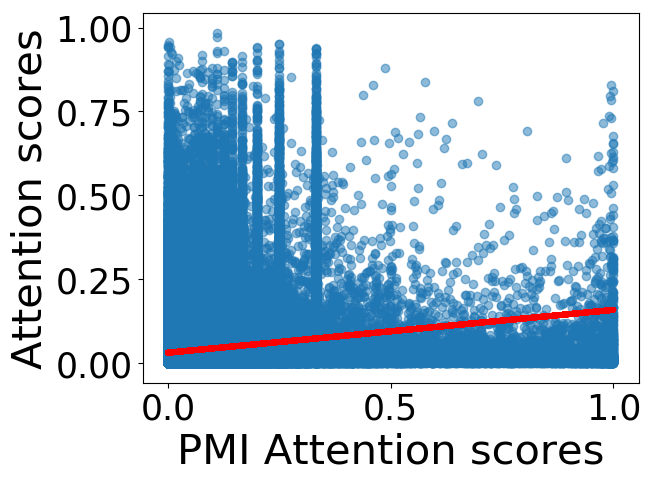}
		\caption{$\rho$=0.153}
	\end{subfigure}
	\begin{subfigure}{0.11\textwidth}
		\includegraphics[width=\textwidth]{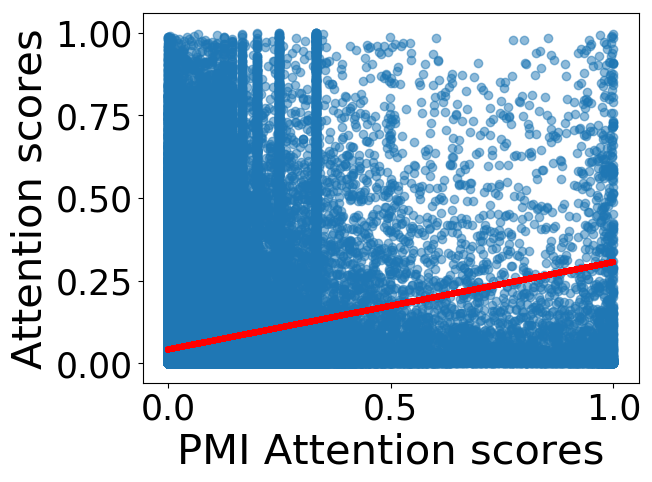}
		\caption{$\rho$=0.215}
	\end{subfigure}
	\begin{subfigure}{0.11\textwidth}
		\includegraphics[width=\textwidth]{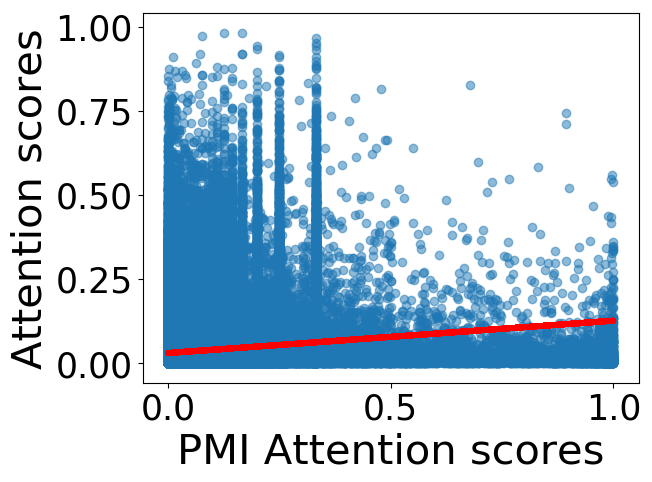}
		\caption{$\rho$=0.171}
	\end{subfigure}
	\begin{subfigure}{0.11\textwidth}
		\includegraphics[width=\textwidth]{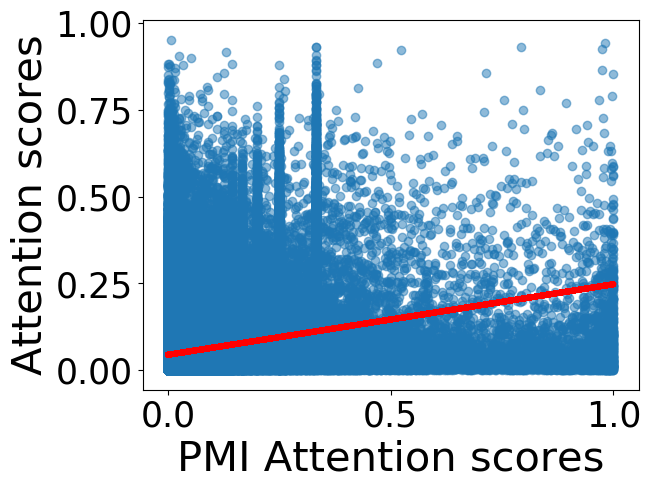}
		\caption{$\rho$=0.254}
	\end{subfigure}
    \vspace{-10pt}
    \caption{Scatter plots of PMI attention scores vs. attention weights learned by various attention mechanisms, showing corresponding Pearson correlation score $\rho$). (a)non-mem + dot, (b)non-mem + metric, (c)mem + dot, (d)mem + metric.}
    \label{fig:pmi-att-vs-att}
    \vspace{-10pt}
\end{figure}

\vspace{-5pt}
\subsubsection{Is our memory metric-based attention helpful?} To answer this question, we evaluate how \emph{MASS}'s performance changed when varying its attention mechanism as follows:
\squishlist
    \item \emph{non-memory + dot product} (\emph{non-mem + dot}): It is the popular \emph{dot attention} introduced in \cite{luong2015effective}.
    \item \emph{non-memory + metric} (\emph{non-mem + metric}): It is our proposed attention with Mahalanobis distance but no external memory.
    \item \emph{memory + dot product} (\emph{mem + dot}):  It is the \emph{dot attention} but exploiting external memory.
    \item \emph{memory + metric} (\emph{mem + metric}): It is our proposed attention mechanism.
\squishend

We do not compare with the \emph{no-attention} case because literature has already proved the effectiveness of the attention mechanism \cite{vaswani2017attention}. Table~\ref{table:attentionComparision} shows the performance of \emph{MASS} under the variations of our proposed attention mechanism. We have some key observations. First, \emph{non-mem + metric} attention outperforms \emph{non-mem + dot} attention with an improvement of 4.9\% on average. Similarly, \emph{mem + metric} attention improves the \emph{mem + dot} attention design by 5.4\% on average. This enhancement comes from different nature of metric space and dot product space. Moreover, these results confirm that metric-based attention designs fit better into our proposed Mahalanobis distance based model. Second, \emph{memory} based attention works better than \emph{non-mem} attention. Particularly, on average, \emph{mem + dot} improves \emph{non-mem + dot} by 2.98\%, and \emph{mem + metric} improves \emph{non-mem + metric} by 3.43\%. Overall, the performance order is \emph{mem + metric} > \emph{non-mem + metric} > \emph{mem + dot} > \emph{non-mem + dot}, which confirms that our proposed attention performs the best and improves 3.43$\sim$8.51\% compared to its variations.


\vspace{-5pt}
\subsubsection{Deep analysis on attention} To further understand how attention mechanisms work, 
we connect attentive scores generated by attention mechanisms with \emph{Pointwise Mutual Information} scores. Given a target song $k$ and a member song $t$, the \emph{PMI} score between them is defined as: $PMI(k, t)=log\frac{P(k, t)}{P(k)\times P(t)}$. Here, \emph{PMI}(k,t) score indicates how likely two songs $k$ and $t$ co-occur together, or how likely a target song $k$ will be added into song $t$'s playlist.

Given a playlist that has a set of $l$ member songs, we measure \emph{PMI} scores between the target song $k$ and each of $l$ member songs. Then, we apply $softmax$ to those \emph{PMI} scores to obtain \emph{PMI attentive scores}. Intuitively, the member song $t$ that has a higher \emph{PMI} score with candidate song $k$ (i.e., co-occurs more with song $k$) will have a higher \emph{PMI attentive score}. We draw scatter plots between \emph{PMI attentive scores} and attentive scores generated by our proposed attention mechanism and its variations. Figure \ref{fig:pmi-att-vs-att} shows the experimental results. We observe that the Pearson correlation $\rho$ between the \emph{PMI attentive scores} and the attentive scores generated by our attention mechanism is the highest (0.254). This result shows that our proposed attention tends to give higher scores to co-occurred songs, which is what we desire. The Pearson correlation results are also consistent with what was reported in Table \ref{table:attentionComparision}.

\begin{figure}[t]
	\centering
	\centering
	\includegraphics[width=0.42\textwidth]{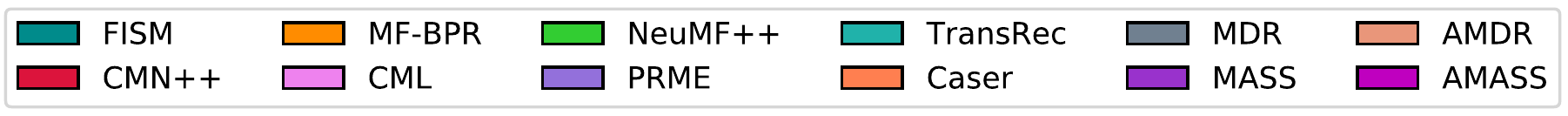}
	\includegraphics[width=0.22\textwidth]{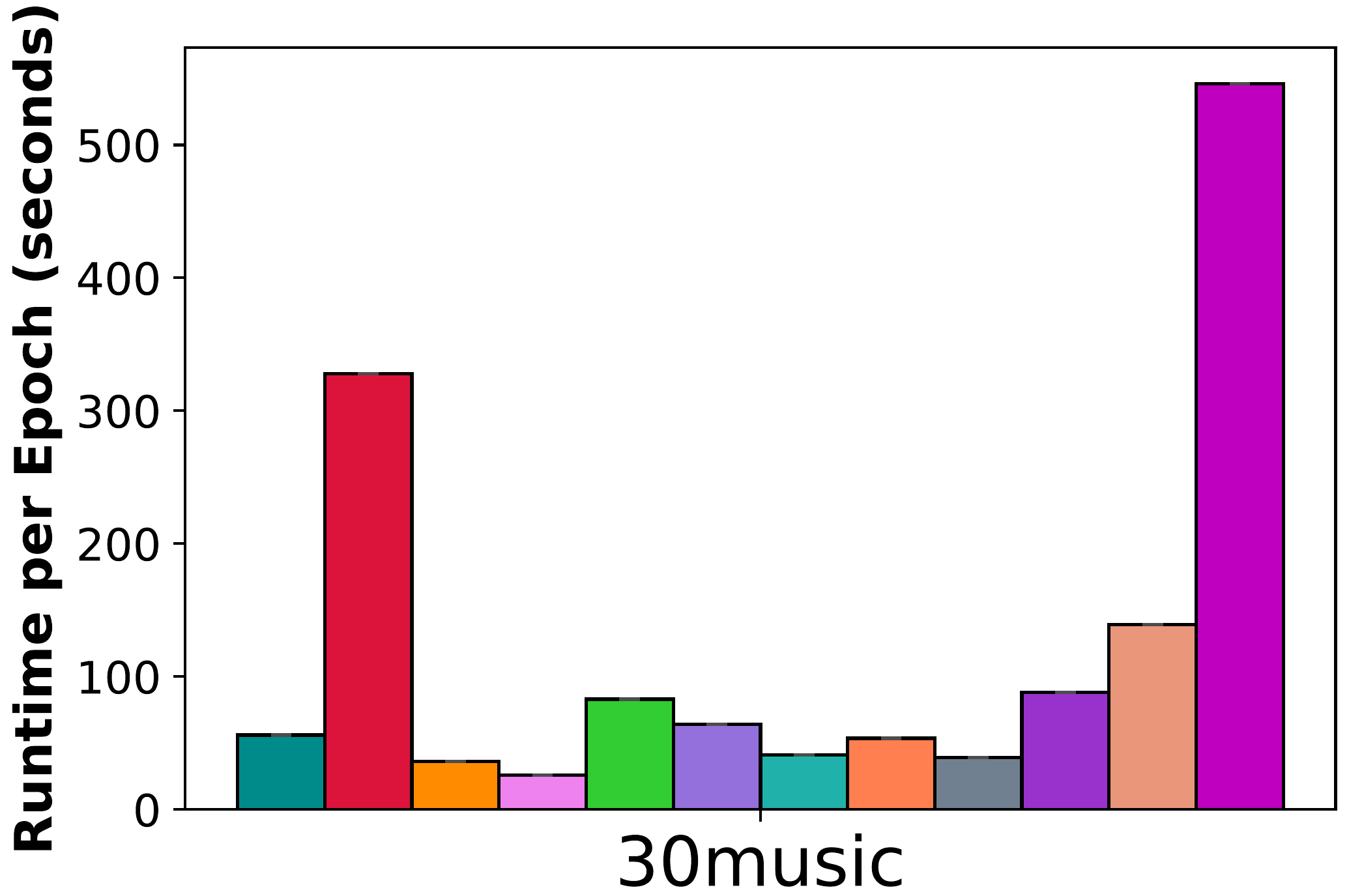}
	\includegraphics[width=0.22\textwidth]{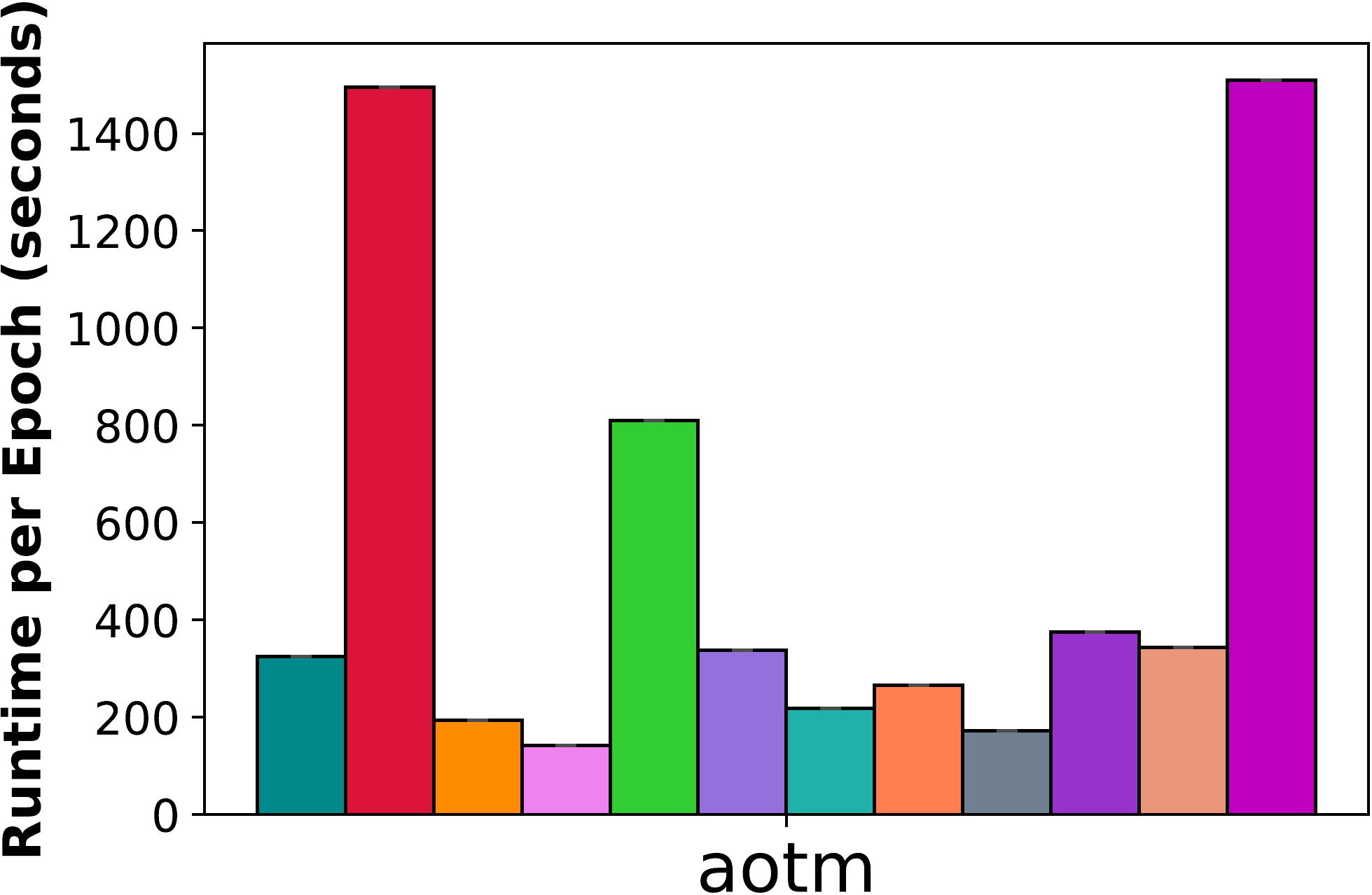}
	\vspace{-10pt}
	\caption{Runtime of all models in 30Music and AOTM.}
	\label{fig:running-time-report}
	\vspace{-15pt}
\end{figure}

\vspace{-5pt}
\subsubsection{Runtime comparison}
To compare model runtimes, we used a Nvidia GeForce GTX 1080 Ti with a batch size of 256 and embedding size of 64. We do not report \emph{MASR} and \emph{AMASR}'s runtimes because their components are pretrained and fixed (i.e., there is no learning process/time). Figure~\ref{fig:running-time-report} shows the runtimes (seconds per epoch) of our models and the baselines for each dataset. \emph{MDR} only took 39 and 173 seconds per epoch in \emph{30Music} and \emph{AOTM}, respectively, while \emph{MASS} took 88 and 375 seconds. \emph{MDR}, one of the fastest models, was also competitive with \emph{CML} and \emph{MF-BPR}.

\section{Conclusion}
In this work, we proposed three novel recommendation approaches based on Mahalanobis distance. Our \emph{MDR} model used Mahalanobis distance to account for both users' preferences and playlists' themes over songs. Our \emph{MASS} model measured attentive similarities between a candidate song and member songs in a target playlist through our proposed memory metric-based attention mechanism. Our \emph{MASR} model combined the capabilities of \emph{MDR} and \emph{MASR}. We also adopted and customized \emph{Adversarial Personalized Ranking} (APR) loss with proposed flexible noise magnitude to further enhance the robustness of our three models. Through extensive experiments against eight baselines in two real-world large-scale \emph{APC} datasets, we showed that our \emph{MASR} improved 20.3\%, and \emph{AMASR} using \emph{APR} loss improved 27.3\% on average over the best baseline. Our runtime experiments also showed that our models were not only competitive, but fast as well.

\section*{Acknowledgment}
This work was supported in part by NSF grant CNS-1755536, Google Faculty Research Award, Microsoft Azure Research Award, AWS Cloud Credits for Research, and Google Cloud. Any opinions, findings and conclusions or recommendations expressed in this material are the author(s) and do not necessarily reflect those of the sponsors.
 
\small
\bibliographystyle{ACM-Reference-Format}
\bibliography{ref}
\end{document}